\documentclass[runningheads,a4paper,11pt,onecolumn]{llncs}
\usepackage[margin=0.8in]{geometry}
\usepackage{graphicx}
\usepackage[utf8]{inputenc}
\RequirePackage{fix-cm}
\usepackage{color}
\usepackage{mathptmx}      
\usepackage{latexsym}
\usepackage{amssymb}
\usepackage{nameref}
\usepackage{amsmath}
\usepackage{hyperref}
\usepackage{mathptmx}      
\usepackage{latexsym}
\usepackage{varioref}
\usepackage{booktabs}  %
\usepackage{url}
\urldef{\mailsa}\path|{alfred.hofmann, ursula.barth, ingrid.haas, frank.holzwarth,|
\urldef{\mailsb}\path|anna.kramer, leonie.kunz, christine.reiss, nicole.sator,|
\urldef{\mailsc}\path|erika.siebert-cole, peter.strasser, lncs}@springer.com|    
\newcommand{\keywords}[1]{\par\addvspace\baselineskip
\noindent\keywordname\enspace\ignorespaces#1}

\begin{document}
\mainmatter  
\title{Fog Computing in Medical Internet-of-Things: Architecture, Implementation, and Applications}
\titlerunning{Chapter in Springer Handbook of Large-Scale Distributed Computing in Smart Healthcare}
%
%
\author{Harishchandra Dubey\textsuperscript{1,3,5}\thanks{\textcolor{blue}{This material is presented to ensure timely dissemination of scholarly and technical work. Copyright and all rights therein are retained by the authors or by the respective copyright holders. The original citation of this book chapter is: H. Dubey , N. Constant, M. Abtahi, A. Monteiro, D. Borthakur, L. Mahler, Y. Sun, Q. Yang, U. Akbar, K. Mankodiya, "Fog Computing in Medical Internet-of-Things: Architecture, Implementation, and Applications", Chapter in Handbook of Large-Scale Distributed Computing in Smart Healthcare (2017), Springer International Publishing AG, S.U. Khan et al. (eds.), Handbook of Large-Scale Distributed Computing in Smart Healthcare, Scalable Computing and Communications, DOI 10.1007/978-3-319-58280-1\_11.}} \and Admir Monteiro\textsuperscript{1,3} \and Nicholas Constant\textsuperscript{1,3} \and Mohammadreza Abtahi\textsuperscript{1,3} \and Debanjan Borthakur\textsuperscript{1,3} \and Leslie Mahler\textsuperscript{2}\and Yan Sun\textsuperscript{1}\and Qing Yang\textsuperscript{1}\and Umer Akbar\textsuperscript{4} \and Kunal Mankodiya\textsuperscript{1,3}
\institute{\textsuperscript{1} Department of Electrical, Computer, and Biomedical Engineering, University of Rhode Island, RI-02881, USA\\
\textsuperscript{2} Department of Communicative Disorders, University of Rhode Island, RI-02881, USA\\
\textsuperscript{3} Wearable Biosensing Lab, University of Rhode Island, RI-02881, USA\\
\textsuperscript{4} Movement Disorders Program, Rhode Island Hospital, RI-02903, USA\\
\textsuperscript{5} Center for Robust Speech Systems, University of Texas at Dallas, TX-75080, USA\\
kunalm@uri.edu\\
}}
\maketitle
\begin{abstract}
In the era when the market segment of Internet of Things (IoT) tops the chart in various business reports, it is apparently envisioned that the field of medicine expects to gain a large benefit from the explosion of wearables and internet-connected sensors that surround us to acquire and communicate unprecedented data on symptoms, medication, food intake, and daily-life activities impacting one's health and wellness. However, IoT-driven healthcare would have to overcome many barriers, such as: 1) There is an increasing demand for data storage on cloud servers where the analysis of the medical big data becomes increasingly complex; 2) The data, when communicated, are vulnerable to security and privacy issues; 3) The communication of the continuously collected data is not only costly but also energy hungry; 4) Operating and maintaining the sensors directly from the cloud servers are non-trial tasks.

This book chapter defined Fog Computing in the context of medical IoT. Conceptually, Fog Computing is a service-oriented intermediate layer in IoT, providing the interfaces between the sensors and cloud servers for facilitating connectivity, data transfer, and queryable local database. The centerpiece of Fog computing is a low-power, intelligent, wireless, embedded computing node that carries out signal conditioning and data analytics on raw data collected from wearables or other medical sensors and offers efficient means to serve telehealth interventions. We implemented and tested an fog computing system using the Intel Edison and Raspberry Pi that allows acquisition, computing, storage and communication of the various medical data such as pathological speech data of individuals with speech disorders, Phonocardiogram (PCG) signal for heart rate estimation, and Electrocardiogram (ECG)-based Q, R, S detection. The book chapter ends with experiments and results showing how fog computing could lessen the obstacles of existing cloud-driven medical IoT solutions and enhance the overall performance of the system in terms of computing intelligence, transmission, storage, configurable, and security. The case studies on various types of physiological data shows that the proposed Fog architecture could be used for signal enhancement, processing and analysis of various types of bio-signals. 

\keywords{Big Data, Body Area Network, Body Sensor Network, Edge Computing, Fog Computing, Medical Cyber-physical Systems, Medical Internet-of-Things, Telecare, Tele-treatment, Wearable Devices.}
\end{abstract}

\section{Introduction}
\begin{figure}[!tb]
\centering
\includegraphics[width=450bp]{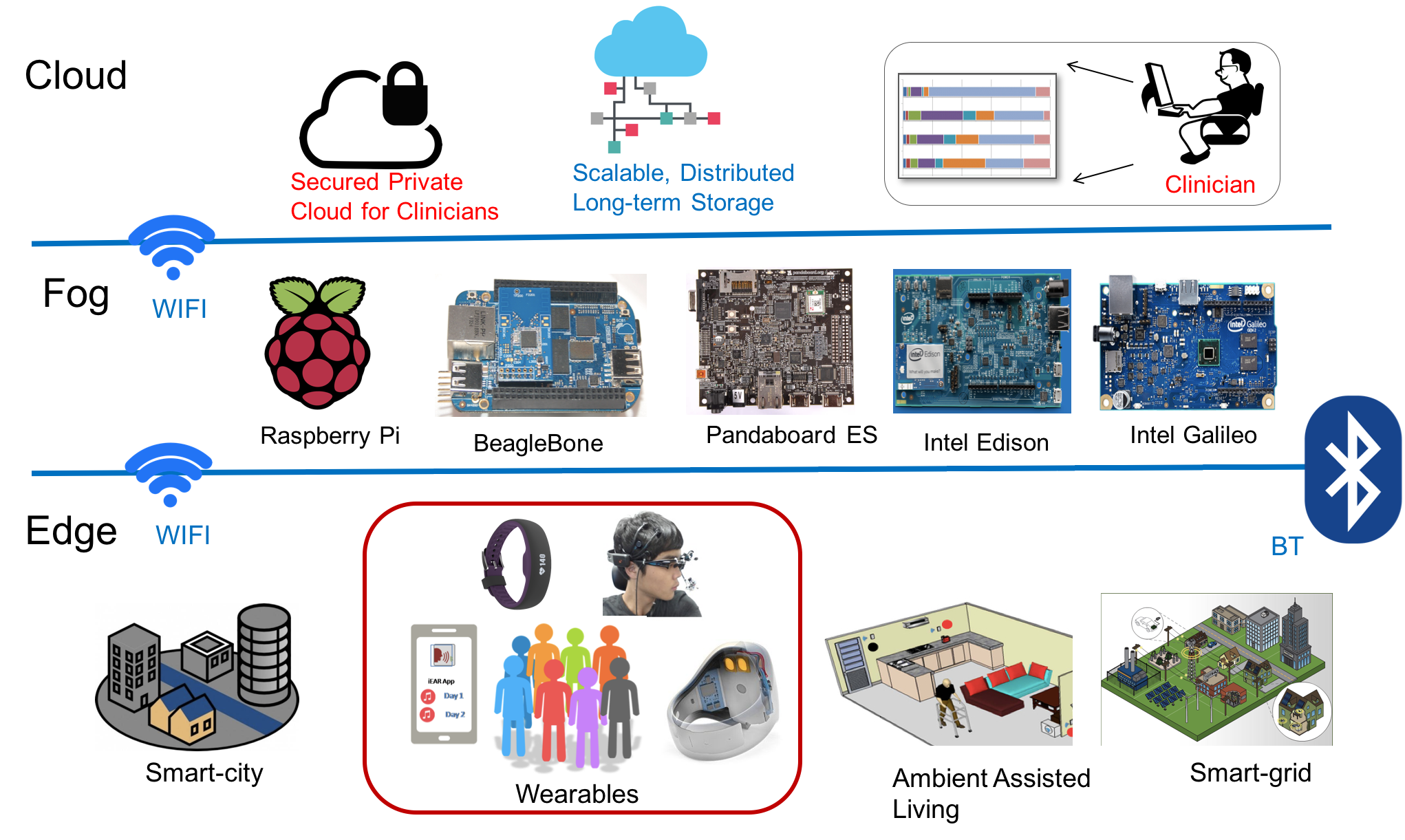}
\caption{Fog computing as an intermediate computing layer between edge devices (wearables) and cloud (backend). The Fog computer enhances the overall efficiency by providing computing near the edge devices. Such frameworks are useful for wearables (employed for healthcare, fitness and wellness tracking), smart-grid,  smart-cities and ambient-assisted living~\emph{etc.}.} 
\label{fig_fog_arch}
\end{figure}
The recent advances in Internet of Things (IoT) and growing use of wearables for the collection of physiological data and bio-signals led to an emergence of new distributed computing paradigms that combined  wearable devices with the medical internet of things for scalable remote tele-treatment and telecare~\cite{dubey2016bigear,dubey2015motor,dubey2016hsum}. Such systems are useful for wellness and fitness monitoring, preliminary diagnosis and long-term tracking of patients with acute disorders. Use of Fog computing reduces the logistics requirements and cut-down the associated medicine and treatment costs (See Figure~\ref{fig_fog_arch}). Fog computing have found emerging applications into other domains such as geo-spatial data associated with various healthcare issues~\cite{barik2016foggis}. 

This book chapter highlights the recent advancements and associated challenges in employing wearable internet of things (wIoT) and body sensor networks (BSNs) for healthcare applications. We present the research conducted in Wearable Biosensing Lab and other research groups at the University of Rhode Island. We developed prototypes using Raspberry Pi and Intel Edison embedded boards and conducted case studies on three healthcare scenarios: (1) Speech Tele-treatment of patients with Parkinson's disease; (2) Electrocardiogram (ECG) monitoring; (3) Phonocardiography (PCG) for heart rate estimation. This book chapter extends the methods and systems published in our earlier conferences papers by adding novel system changes and algorithms for robust estimation of clinical features.

This chapter made the following contributions to the area of \textit{Fog Computing for Medical Internet-of Things}:
\begin{itemize}
\item \textbf{Fog Hardware:} Intel Edison and Raspberry Pi were leveraged to formulate two prototype architectures. Both of the architectures can be used for each of the three case-studies mentioned above. 

\item \textbf{Edge Computing of Clinical Features:} The Fog devices executed a variety of algorithms to extract clinical features and performed primary diagnosis using data collected from wearable sensors;

\item \textbf{Interoperability:} We designed frontend apps for body sensor network such as android app for smartwatch~\cite{hermes}, PPG wrist-band, and backend cloud infrastructure for long-term storage. In addition, transfer, communication, authentication, storage and execution procedures of data were implemented in the Fog computer.

\item \textbf{Security:} In order to ensure security and data privacy, we built an encrypted server that handles user authentication and associated privileges. The rule-based authentication scheme is also a novel contribution of this chapter where only the individuals with privileges (such as clinicians) could access the associated data from the patients. 

\item \textbf{Case Study on Fog Computing for Medical IoT-based Tele-treatment and Monitoring:} We conducted three case studies: (1) Speech Tele-treatment of patients with Parkinson's disease; (2) Electrocardiogram (ECG) monitoring; (3) Phonocardiography (PCG) for heart rate estimation. Even if we conducted validation experiments on only three types of healthcare data, the proposed Fog architecture could be used for analysis of other bio-signals and healthcare data as well. 

\item \textbf{Android API for Wearable (Smartwatch-based) Audio Data Collection} The EchoWear app that was introduced in~\cite{dubey2015echowear} is used in proposed architecture for collecting the audio data from wearables. We have released the library to public at: \textit{https://github.com/harishdubey123/wbl-echowear}
\end{itemize}

\section{Related~\& Background Works}
In this section, we present recent emergence of wearables and fog computing for enhancement processing of physiological data for healthcare applications. 
\subsection{Wearable Big Data in Healthcare}
The medical data is collected by the intelligent edge devices such as wearables, wrist-bands, smartwatches, smart textiles~\emph{etc.}. The intelligence refers to knowledge of analytics, devices, clinical application and the consumer behavior. Such smart data is structured, homogeneous and meaningful with negligible amount of noise and meta-data~\cite{siemens}. The big data and quiet recently smart data trend had revolutionized the biomedical and healthcare domain. With increasing use of wireless and wearable body sensor networks (BSNs), the amount of data aggregated by edge devices and synced to the cloud is growing at enormous rate~\cite{kayyali2013big}. The pharmaceutical companies are leveraging deep learning and data analytics on their huge medical databases. These databases are results of digitization of patient’s medical records. The data obtained from patient’s health records, clinical trials and insurance programs provided an opportunity for data mining. Such databases are heterogeneous, unstructured, scalable and contain significant amount of noise and meta-data. The noise and meta-data have low or no useful information. Cleaning and structuring the real-world data is another challenge in processing medical big data. In recent years, the big data trend had transformed the healthcare, wellness and fitness industry. Adding value and innovation in data processing chain could help patients and healthcare stakeholders accomplish the treatment goals in lower cost with reduced logistic requirements~\cite{kayyali2013big}. Authors in~\cite{panahiazar2014empowering} presented the smart data as a result of using semantic web and data analytics on structured collection of big data. Smart data attempts to provide a superior avenue for better decision and inexpensive processing for person-centered medicine and healthcare. The medical data such as diagnostic images, genetic test results and biometric information are getting generated at large scale. Such data has not just the high volume but also a wide variety and different velocity. It necessitates the novel ways for storing, managing, retrieving and processing such data. The smart medical data demand development of novel scalable big data architecture and applicable algorithms for intelligent data analytics. Authors also underlined the challenges in semantic-rich data processing for intelligent inference on practical use cases~\cite{panahiazar2014empowering}.

\begin{figure*}[!t]
\centering
\includegraphics[width=470pt]{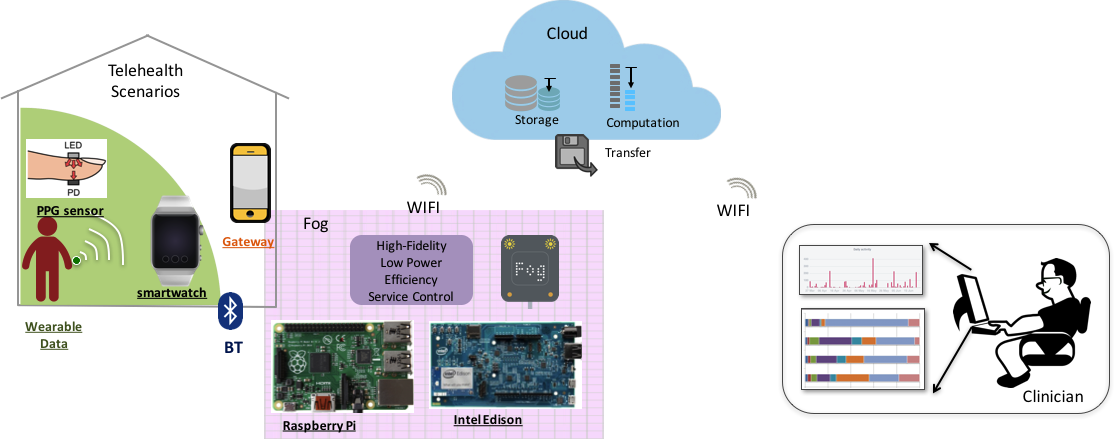}
\caption{The conceptual overview of the proposed Fog architecture that assisted Medical Internet of Things framework in tele-treatment scenarios.}
\label{fig_arch}
\end{figure*}
\begin{table*}[!tb]
\small
\centering
\caption{A comparison between Fog computing and cloud computing [adopted from~\cite{rr40_py}.}
\begin{tabular}{ c c c c}
\toprule
\textbf{Criterion} & \shortstack{\textbf{Fog nodes close to}\\\textbf{ IoT devices}} & \shortstack{\textbf{Fog Aggregation}\\\textbf{Nodes}} &\shortstack{\textbf{Cloud}\\\textbf{Computing}} \\
\hline
\shortstack{Response\\time} & \shortstack{Milliseconds to\\sub-second} & \shortstack{Seconds to\\minutes} & \shortstack{Minutes,\\days, weeks} \\
\hline
\shortstack{Application\\examples} & \shortstack{Telemedicine\\and training} & \shortstack{Visualization\\simple analytics} & \shortstack{Big data\\analytics \\
Graphical\\dashboards}\\
\hline
\shortstack{How long IoT\\data is stored} & Transient&\shortstack{Short duration:\\perhaps hours, days\\or weeks}&\shortstack{Months or\\ years}\\
\hline
\shortstack{Geographic\\coverage}&\shortstack{Very local:\\for example, one\\city block}&Regional&Global\\
\bottomrule
\label{table1}
\end{tabular}
\end{table*}
\begin{figure}[!t]
\centering
\includegraphics[width=250bp]{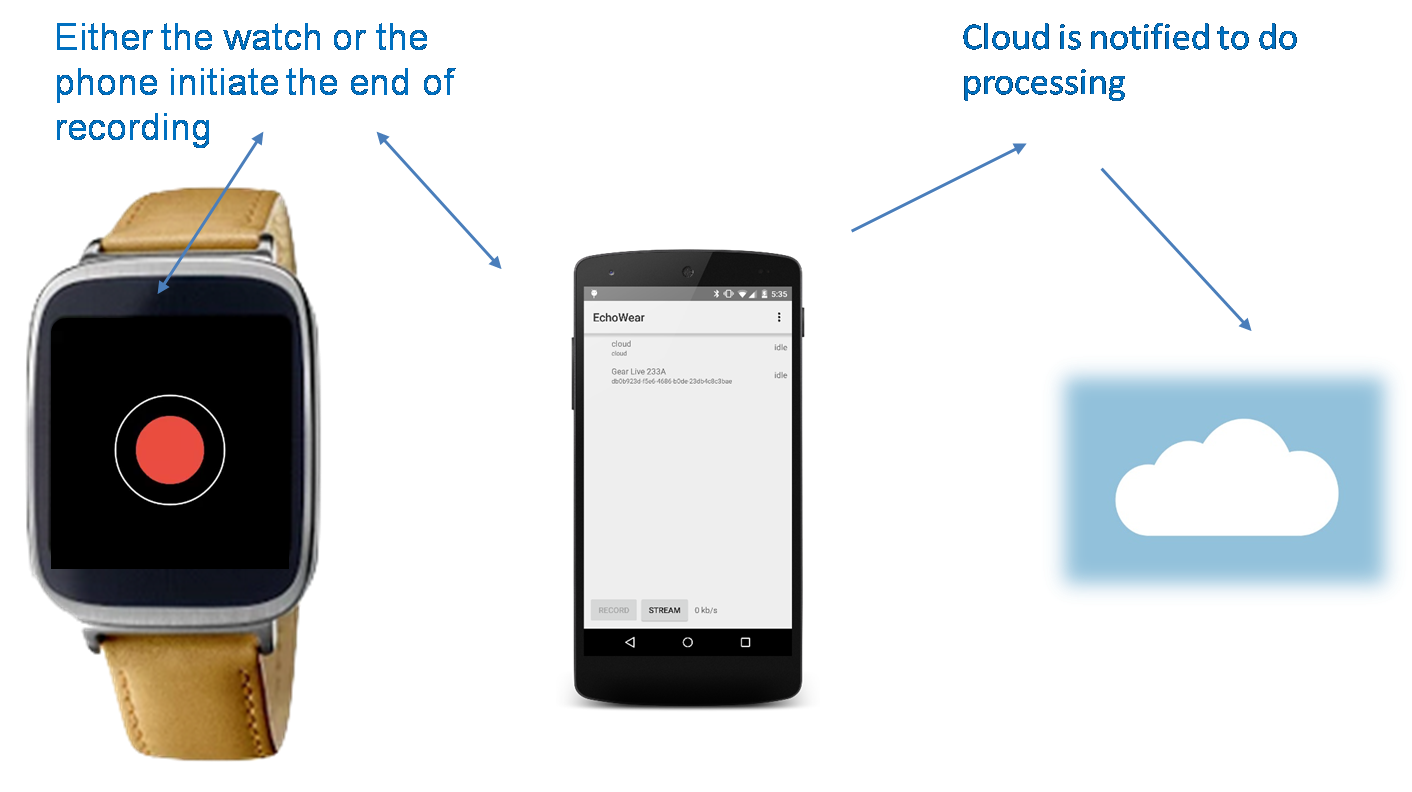}
\caption{The flow of information and control between three main components of the medical IoT system for smartwatch-based speech treatment~\cite{dubey2015echowear}. The smartwatch is triggered by the patients with Parkinson's disease. At fixed timings set by patients, caregivers or their speech-language pathologist (SLPs), the tablet triggers the recording of speech data. The smartwatch interacts with the tablet via Bluetooth. Once tablet gets the data from smartwatch, it send to the Fog devices that process the clinical speech. Finally, the features were sent to the cloud from where those could be queried by clinicians for long-term comparative study. SLPs use the final features for designing customized speech exercises and treatment regime in accordance with patient's communications deficits.} 
\label{fig_pb}
\end{figure}
\begin{figure*}[!t]
\centering
\includegraphics[width=450pt]{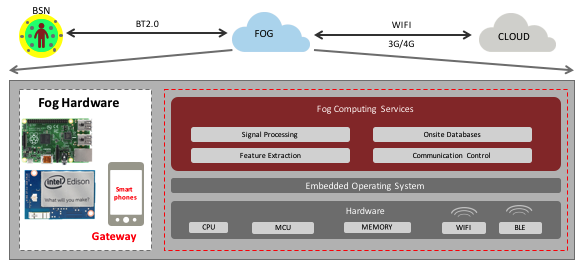}
\caption{The proposed Fog architecture that acquired the data from body sensor networks (BSNs) through smartphone/tablet gateways. It has two choices for fog computers: Intel Edison and Raspberry Pi. The extraction of clinical features was done locally on fog device that was kept in patient's home (or near the patient in care-homes). Finally, the extracted information from bio-signals was uploaded to the secured cloud backend from where it could be accessed by clinicians. The proposed Fog architecture consists of four modules, namely BSNs (e.g. smartwatch), gateways (e.g. smartphone/tablets), fog devices (Intel Edison/Raspberry Pi) and cloud backend.}
\label{fig2}
\end{figure*}
\subsection{Speech Treatments of Patients with Parkinson's Disease}
The patients with Parkinson's Disease (PD) have their own unique set of speech deficits. We developed EchoWear~\cite{dubey2015echowear} as a technology front-end for monitoring the speech from PD patients using smartwatch. The speech-language pathologists (SLPs) had access to such as system for remote monitoring of their patients. The rising cost of healthcare, the increase in elderly population, and the prevalence of chronic diseases around the world urgently demand the transformation of healthcare from a hospital-centered system to a person-centered environment, with a focus on patient's disease management as well as their wellbeing. 

Speech Disorders affected approximately 7.5 million people in US~\cite{nih}. Dysarthria (caused by Parkinson's disease or other speech disorders) refers to motor speech disorder resulting from impairments in human speech production system. The speech production system consists of the lips, tongue, vocal folds, and/or diaphragm. Depending on the part of nervous system that is affected, there are various types of dysarthria. The patients with dysarthria posses specific speech characteristics such as difficult to understand speech, limited movement in lips, tongue and jaw, abnormal pitch and rhythm. It also includes poor voice quality, for instance, hoarse, breathy or nasal voice. Dysarthria results from neural dysfunction. It might happen at birth (cerebral palsy) or developed later in person’s life. It can be due to variety of ailments in the nervous system, such as Motor neuron diseases, Alzheimer’s disease, Cerebral Palsy (CP), Huntington’s disease, Multiple Sclerosis, Parkinson’s disease (PD), Traumatic brain injury (TBI), Mental health issues, Stroke, Progressive neurological conditions, Cancer of the head, neck and throat (including laryngectomy). The patients with dysarthria are subjectively evaluated by the speech-language pathologist (SLP) who identifies the speech difficulties and decide the type and severity of the communication deficit~\cite{asha}. 

Authors in~\cite{sapir2008speech} compared the perceived loudness of speech and self-perception of speech in patients with idiopathic Parkinson's disease (PD) with healthy controls. Thirty patients with PD and fourteen healthy controls participated in the research survey. Various speech tasks were performed and nine speech and voice characteristics were used for evaluation. Results showed that the patients with PD had significant reduction in loudness as compared to healthy controls during various speech tasks. These results furnished additional information on speech characteristics of patients with PD that might be
useful for effective speech treatment of such population~\cite{sapir2008speech}. Authors in~\cite{j2000voice} studied the acoustic characteristics of voice in patients with PD. Thirty patients with early stage PD and thirty patients with later stage PD were compared with thirty healthy controls for acoustic characteristics of the voice. The speech task included sustained /a/ and one minute monologue. The voice of patients with early as well as later stage PD were found to have reduced loudness, limited loudness and pitch variability, breathiness, and harshness. In general, the voice of patients with PD had lower mean intensity levels and reduced maximum phonational frequency range as compared to healthy controls~\cite{j2000voice}.

Authors in~\cite{spielman2011intensive} studied and evaluated the voice and speech quality in patients with and without deep brain stimulation of the subthalamic nucleus (STN-DBS) before and after LSVT LOUD therapy. The goal of the study was to do a comparative study of improvement in surgical patients as compared to the non-surgical ones. Results showed that the LSVT LOUD is recommended for voice and speech treatment of patients with PD following STN-DBS surgery. Authors in~\cite{gamboa1997acoustic} performed acoustic analysis of voice from 41 patients with PD and healthy controls. The speech exercises included in the study were the sustained /a/ for two seconds and reading sentences. The acoustic measures for quantifying the speech quality were fundamental frequency, perturbation in fundamental frequency, shimmer, and harmonic to noise ratio of the sustained /a/, phonation range, dynamic range, and maximum phonation time. Authors concluded that the patients with PD had higher jitter, lower harmonics to noise ratio, lower frequency and intensity variability, lower phonation range, the presence of low voice intensity, mono pitch, voice arrests, and struggle irrespective of the severity of the PD symptoms.

People suffering from Parkinson's disease experience speech production difficulty associated with Dysarthria. Dysarthria is characterized by monotony of pitch, reduced loudness, an irregular rate of speech and, imprecise consonants and changes in voice quality~\cite{lansford2014vowel}. Speech-language pathologists do the evaluation, diagnosis and treat communication disorders. Literature suggests that Lee Silverman Voice Treatment (LSVT) has been most efficient behavioral treatment for voice and speech disorders in Parkinson's disease. Telehealth monitoring is very effective for the speech-language pathology, and smart devices like EchoWear~\cite{dubey2015echowear} can be of much use in such situations. Several cues indicate the relationship of dysarthria and  acoustic features. Some of them are ,

\begin{enumerate}
\item Shallower F2 trajectories in male speakers with dysarthria is observed in~\cite{deb2_kent1999acoustic}.
 \item Vowel space area was found to be reduced relative to healthy controls for male speakers with amyotrophic lateral sclerosis~\cite{deb2_kent1999acoustic}.
\item Shimmers as described in~\cite{dubey2015multi} as a measure of variation in amplitude of the speech and it is an important speech quality metric for people with speech disorders.
\item Like shimmers, Jitters (pitch variations) and loudness and sharpness of the speech signal can be used as a cue for speech disorders~\cite{dubey2015multi}.
\item In ataxic dysarthria ,patients can produce distorted vowels and excess variation in loudness, so speech prosody and acoustic analysis are of much use.
\item Multi dimensional voice analysis as stated in~\cite{deb2_kent1999acoustic} plays an important role in motor speech disorder diagnosis and analysis. Parameters that can effectively used are relative perturbation (RAP), pitch perturbation quotient (PPQ),fundamental frequency variation (vF0), shimmer in dB(ShdB), shimmer percent (Shim), peak amplitude variation (vam) and amplitude tremor intensity index(ATRI).
\item Shrinking of the F0 range as well as vowel space are observed in dysarthria speech. Moreover, from the comparison of F0 range and vowel formant frequencies, it is suggested that speech effort to produce wider F0 range can influence vowel quality as well.	
\end{enumerate}
EchoWear~\cite{dubey2015echowear} is a smartwatch technology for voice and speech treatments of patients with Parkinson's disease. Considering the difficulties associated with the patients in following prescribed exercise regimes outside the clinic, this device remotely monitors speech and voice exercise as prescribed by speech-language pathologists. The speech quality metrics used in EchoWear presently as stated in~\cite{dubey2015echowear} were average loudness level and average fundamental frequency (F0). Features were derived from the short-term speech spectrum of a speech signal. To find the fundamental frequency, EchoWear uses SWIPE pitch estimator, whereas other methods such as cepstral analysis and autocorrelation methods are also extensively used for estimation of the pitch. The software Praat is designed for visualizing the spectrum of a speech signal for analysis. Fundamental frequency (F0) variability is associated with the PD speech. There is a decrease  variation in pitch , i.e. Fundamental frequency associated with PD speech. 

\begin{figure*}[!h]
\centering
\includegraphics[width=450pt]{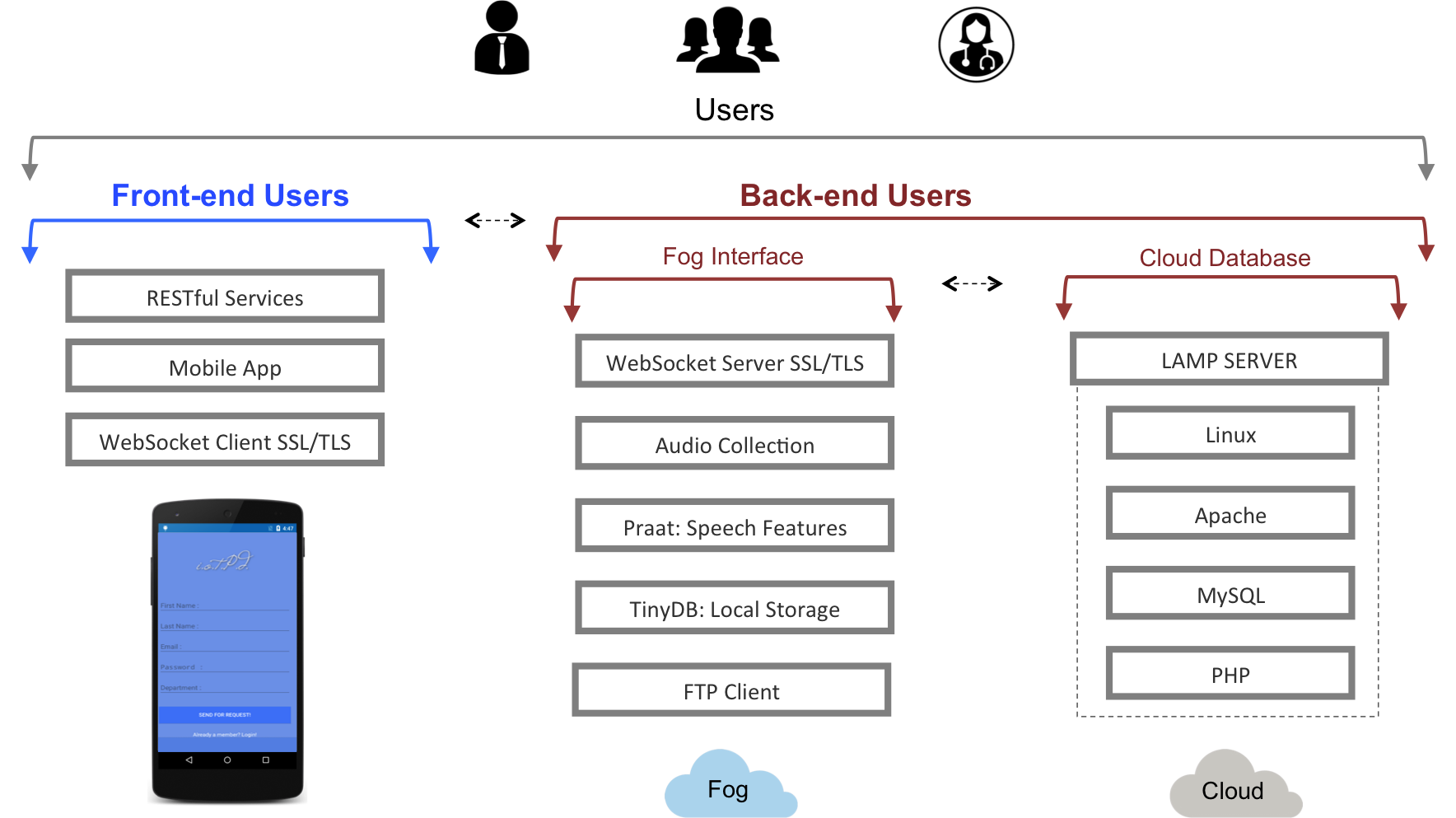}
\caption{Overall architecture of the proposed Fog architecture  in the context of frontend and backend services. It shows the information flow from patients to clinicians through the modular architecture.}
\label{fig4}
\end{figure*}
\begin{table*}[!t]
\small
\centering
\caption{List of speech exercises performed by the patients with Parkinson's disease.}
\begin{tabular}{c c c}
\hline
Task & Exercise Name& Description\\
\hline
$t_{1}$ & Vowel Prolongation & Sustain the vowel /a/ for as long as possible for three repetitions.\\
\hline
$t_{2}$ & High Pitch & Start saying /a/ at talking pitch and then go up and hold for 5 seconds (three repetitions).\\
\hline
$t_{3}$ & Low Pitch & Start saying /a/ at talking pitch and then go down and hold for 5 seconds (three repetitions).\\
\hline
$t_{4}$ & Read Sentence & Read 'The boot on top is packed to keep'\\
\hline
$t_{5}$ & Read Passage & Read the 'farm' passage.\\
\hline
$t_{6}$& Functional Speech Task & Read a set of customized sentences.\\
\hline
$t_{7}$ & Monologue & Explain happiest day of your life.\\
\hline
\label{table_speech_exercises}
\end{tabular}
\end{table*}
\section{Proposed Fog Architecture}
In this section, we describe the implementation of the proposed Fog architecture. Figure~\ref{fig4} shows the overall architecture of proposed system in the context of frontend and backend services. It shows the information
flow from the patients to SLPs through the communication and processing interfaces. Instead of layers, we describe the implementation using three modules namely, 1) Fog device; 2) Backend Cloud Database; and 3) Frontend App Services. These three modules gave a convenient representation for describing the multi-user model of the proposed Fog architecture.
\subsection{Fog Computing Device}
\label{sec:fog}
To transfer the audio file/other data file from a patient, we used socket streaming using TCP wrapped in  Secure Socket Layer/ Transmission Layer Security (SSL/TLS) sockets to ensure the secure transmission. Sockets provide communication framework for devices using different protocols such as TCP and UDP that could then be wrapped in secured sockets. Next, we describe these protocols and their usage in the proposed architecture.
\begin{itemize}
\item \textbf{Transmission Control Protocol (TCP)} is a networking protocol that allows guaranteed and reliable delivery of files. It is a connection-oriented and bi-directional protocol. In other words, both devices could send and receive files using this protocol. Each point of the connection involved Internet Protocol (IP) address and a port number so the connection could be made with a specific device. Furthermore, we wrapped the TCP sockets in SSL Sockets for ensuring the security and privacy of the data collected from the users/patients.

\item \textbf{Secure Sockets Layer (SSL)} is a network communication protocol that allows encrypted authentication for network sockets from the server and client sides. To implement it in the proposed Fog architecture, we used two python modules, namely \textit{SSL} and \textit{socket}. To create the certifications for the server and client, we also used the command line program called \textit{OpenSSL}~\cite{rr32_py}. \textit{OpenSSL} is an open-source project that provides a robust, commercial-grade, and full-featured toolkit for the Transport Layer Security (TLS) and Secure Sockets Layer (SSL) protocols.
\end{itemize}
Once all the SSL certification keys were built for client (the Android gateway devices/wearables) and server (Fog computer e.g., Intel Edison or Raspberry Pi), we ran the secure sockets on the server and continuously listened for a connection for file transfer. We renamed the file with date and time stamps before it could be used for further processing. As soon as the audio file was completely transferred, the connection was closed and the processing began. We used the python based \textit{Praat} and \textit{Christian's Library} described for processing and analysis of audio data. For other healthcare data such as Phonocardiography (PCG) data and Electrocardiogram
(ECG) data~\emph{etc.}, we implemented the associated methods using Python, C and GNU Octave.

\begin{figure*}[!t]
\centering
\includegraphics[width=480pt]{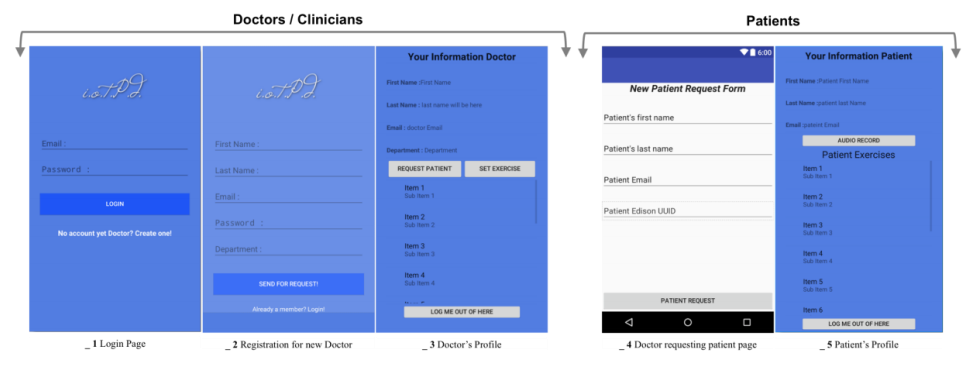}
\caption{The interface view of the IoT PD Android app for frontend users such as clinicians, caregivers and patients. Different categories of users have different privileges. For example, a patient can register with the app only upon receiving the clinician's approval.}
\label{fig6}
\end{figure*}
\subsection{Frontend App Services}
For the frontend users, including patients and clinicians (SLPs), we designed Android applications and web applications that could be used to log-into the system and access clinical features. Also, front-end apps were running on wearable devices are facilitating the data collection. Our app,~\textit{IoT PD}, took advantage of the REST protocol. We used REST protocol for simplicity of implementation. For every REST request of data information gathering, we returned a JSON (JavaScript Object Notation), a format of data-interchange between programs~\cite{json}. The~\textit{IoT PD} app is based on software engine Hermes. We open-source the URI library for audio data collection from wearable devices. 

The app allowed access to two categories of users as shown in Figure~\ref{fig6}. Both the patients and healthcare providers were allowed to login and view their profile; however their profiles were different, only the clinicians could give permission to their patients for app registration. Further, the physician could setup personalized notifications for their patients. For example, the physician could schedule a personalized exercise regime for a given patient so that their speech functions could be enhanced. On the other hand, patients could only view their information and visual data.
\subsection{Backend Cloud Database}
To support the centralized storage of clinical features and analytics, we implemented a backend cloud database using \textit{PHP} and \textit{MySQL}. Firstly, we set up a \textit{Linux, Apache, MySQL, PHP (LAMP) server}, an open-source web platform for development on \textit{Linux} systems using \textit{Apache} for
web servicing, \textit{MySQL} as database system for management and storage, and \textit{PHP} as the language for server interaction with applications~\cite{rr33_sobell2013practical}. The main component of the backend was the relational database development. We designed a database revolving around the users and Fog computers that could easily engage with the database. It created three tables that were used for the users (patients and healthcare providers such as clinicians). The fourth table was created for the information extracted from the patient's data. The extracted features obtained from the Fog computer were entered in the data table.
\subsection{Pathological Speech Data Collection}
Earlier, we described our implementation of EchoWear that was used in an in-clinic validation study on six patients with Parkinson's disease (PD). We received an approval (no: 682871-2) of the University of Rhode Island's Institutional Review Board to conduct human studies involving the presented technologies including~\textit{IoT PD} and proposed Fog architecture. First, the six patients were given an intensive voice training in the clinic by Leslie Mahler, a speech-language pathologist, who also prescribed home speech tasks for each patient. Patients were given a home kit consisting of a smartwatch, a companion tablet and charging accessories. Patients were recommended to wear the smartwatch during the day. Patients chose their preferred timings for speech exercise. A tactile vibration of the smartwatch was used as a notification method to remind the patients to perform speech exercises. The \textit{IoT PD} app took the timings to set the notifications accordingly. Home exercise regime had six speech tasks. The six speech tasks assigned to patients with PD are given in Table~\ref{table_speech_exercises}. Speech-language pathologists (SLPs) use extensive number of speech parameters in their diagnosis. We skip the clinical details of prescription as it is out-of-the-scope of this book chapter. 
\subsection{Dynamic Time Warping}
Dynamic time warping (DTW) is an algorithm for finding
similar patterns in a time-series data. DTW has been used for
time-series mining for a variety of applications such as
business, finance, single word recognition, walking pattern
detection, and analysis of ECG signals. Usually, we use
Euclidean distance to measure the distance between two
points. For example, consider two vectors , $x= [x_{1}, x_{2},...,x_{n}]$ and $y= [y_{1}, y_{2},...,y_{n}]$
\begin{equation}
d(x,y)= \sqrt{(x_{1} - y_{1})^2 + (x_{2} - y_{2})^2+ (\cdot x_{n} - y_{n})^2}
\end{equation}
Euclidean distance works well in many areas. But for some special case where two similar and out-of-phase series are to be compared, Euclidean distance fails to detect similarity. For example, consider two time series A = [1,1,1,2,8,1] and B
[1,1,1,8,2,1], the Euclidean distance between them is $\sqrt{72}$. Thus, DTW is an effective algorithm that can detect the similarity between two series regardless of different length, and/or phase difference. The example vectors are similar but the similarity could not be inferred by Euclidean distance metric while DTW can detect the similarity easily. DTW is based on the idea of dynamic programming (DP). It builds an adjacency matrix then finds the shortest path across it. DTW is more effective than
Euclidean distance for many applications~\cite{dtw21_ding2008querying} such as gesture recognition~\cite{dtw22_gavrila1995towards}, fingerprint verification~\cite{dtw23_kovacs2000fingerprint}, and speech
processing~\cite{dtw24_myers1980performance}.
\begin{figure*}[!t]
\centering
\includegraphics[width=490pt]{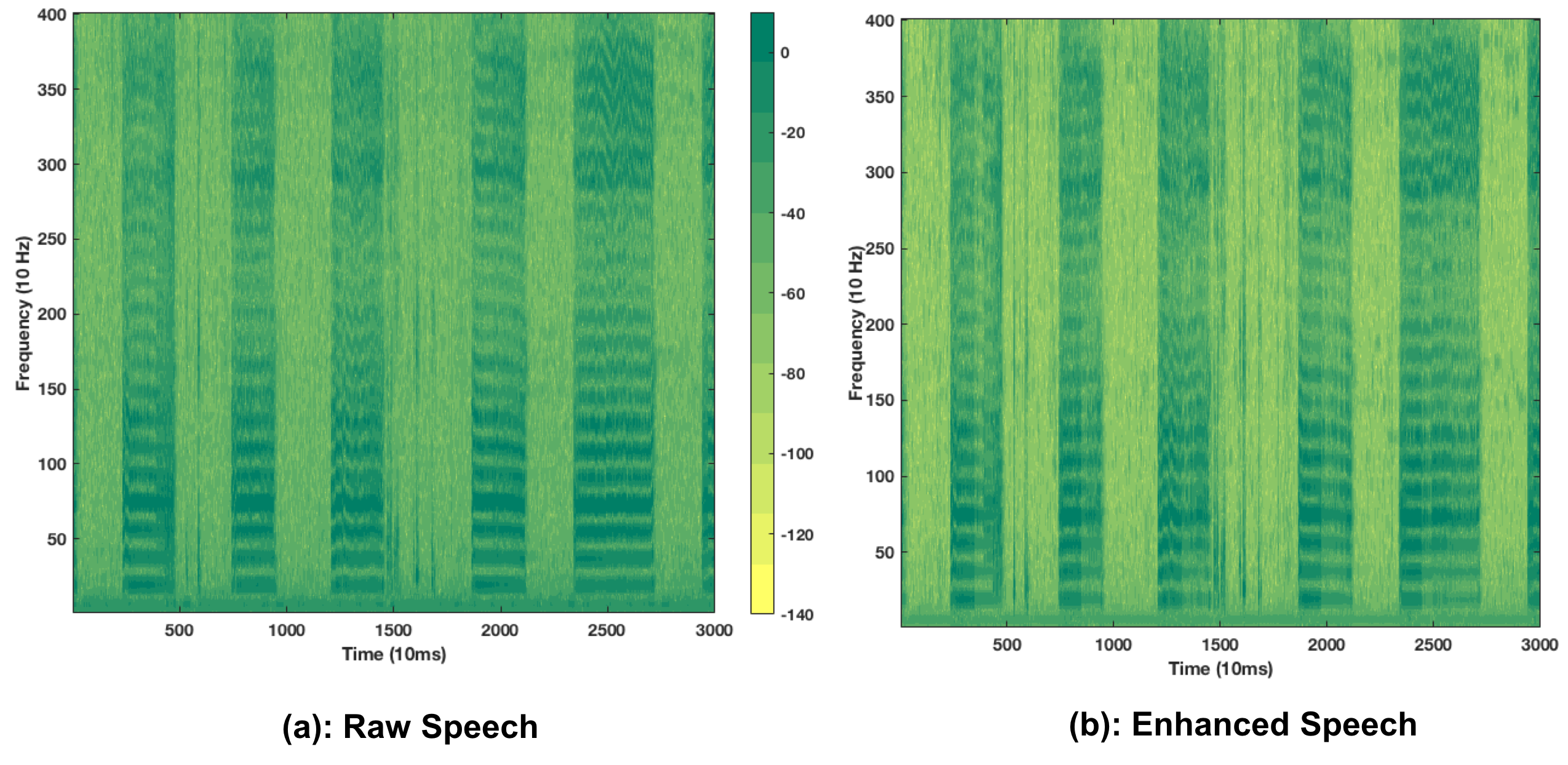}
\caption{(a)Spectrogram of acquired speech signal. The frequency sampling rate is 8000 Hz. Time-windows of 25 ms with 10 ms skip-rate were used. (b) Spectrogram of enhanced speech signal.}
\label{fig_orig_enh_spec}
\end{figure*}
\begin{figure*}[!t]
\centering
\includegraphics[width=350pt]{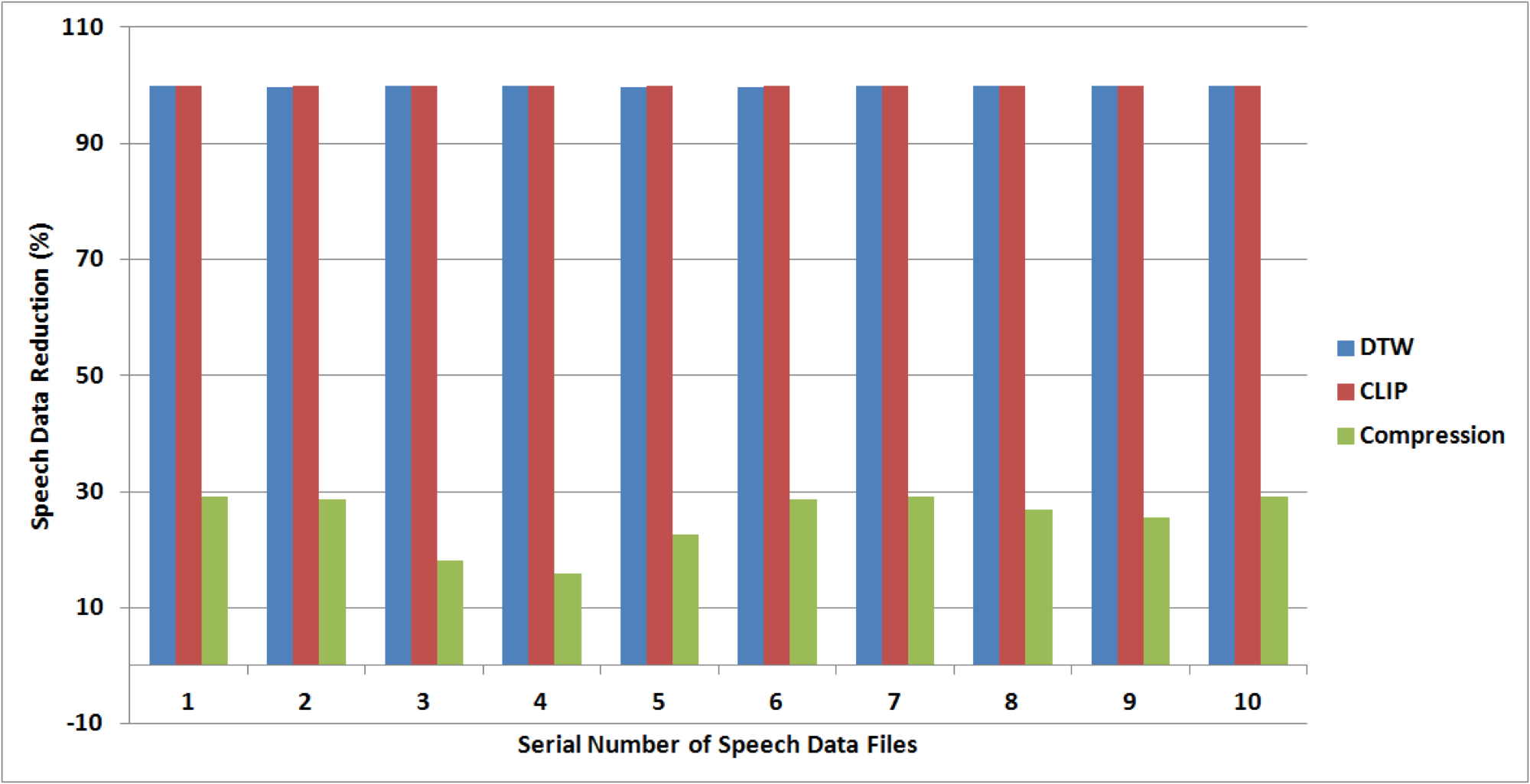}
\caption{Bar chart depicting the data reduction achieved by using dynamic time warping (DTW), clinical speech processing (CLIP) and GNU zip compression on
ten sample speech files collected from in-home trails of patients with Parkinson's disease (PD).}
\label{fig_data_reduce}
\end{figure*}
\section{Case Studies using Proposed Fog Architecture}
\subsection{Case Study I: Speech Tele-treatment of Patients with Parkinson's Disease}

A variety of acoustic and spectral features were derived from the speech content of audio file acquired by wearables. In proposed Fog architecture, noise reduction, automated trimming, and feature extraction were done on the Fog device. In our earlier studies~\cite{dubey2015echowear,dubey2015fogdata,constant2017fog,monteiro2016fit}, trimming was done manually by human annotator and feature extraction was done in the cloud. In addition, there were no noise reduction done in previous studies~\cite{dubey2015echowear,dubey2015fogdata}. The Fog computer syncs the extracted features and preliminary diagnosis back in the secured cloud backend. Fog was employed for in-home speech treatment of patients with Parkinson's disease. The pathological features were later extracted from the audio signal. Figure~\ref{fig_speech_proc} shows the block diagram of pathological speech processing module. In our earlier studies, we computed features from the controlled clinical environment and performed Fog device trails in lab scenarios~\cite{dubey2015fogdata}. This paper explored the in-home field application. In-clinic speech data was obtained in quiet scenarios with negligible background noise. On the other hand, data from in-home trials had huge amounts of time-varying non-stationary noise. It necessitated the use of robust algorithms for noise reduction before extracting the pathological features. In addition to previously studied features such as loudness and fundamental frequency, we developed more features for accurate quantification of abnormalities in patient's vocalization. The new features are jitter, frequency modulation, speech rate and sensory pleasantness. In our previous studies, we use just three speech exercises (tasks $t_{1}$, $t_{2}$ and $t_{3}$) for analysis of algorithms. In this paper, we incorporated all six speech exercises. The execution was done in real-time in patient's home unlike pilot data used in our previous studies~\cite{dubey2015fogdata,monteiro2016fit}. Thus, Fog speech processing module is an advancement over earlier studies in~\cite{dubey2015echowear,dubey2015fogdata,monteiro2016fit}. 

The audio data was acquired and stored in \textit{wav} format. Using perceptual audio coding such as \textit{mp3} would have saved transmission power, storage and execution time as the size of mp3 coded speech data is lower than corresponding wav format. The reason for not using mp3 or other advanced audio codecs is to avoid loss of information. Perceptual audio codes such as mp3 are lossy compression scheme that removes frequency bands that are not perceptually important. Such codecs have worked well for music and audio streaming. However, in pathological speech analysis, patients have very acute vocalizations such as nasal voice, hypernasal voice, mildly slurred speech, monotone voice~\emph{etc.}. Clinicians do not recommend lossy coding for speech data as it can cause confusion in diagnosis, monitoring, and evaluation of pathological voice. Since we use the unicast transmission from BSNs to fog computers, we employed Transmission Control Protocol (TCP). The data have to be received in the same order as sent by BSNs. We did not use User Datagram Protocol (UDP) that is more popular for audio/video streaming as UDP does not guarantee receipt of packets. For videos/audios that are perceptually encoded and decoded, small losses lead to temporary degradation in received audio/video. We do not have that luxury in pathological speech or PCG data that have to be guaranteed delivery even if delayed and/or have to be re-transmitted. The pathological data was saved as mono-channel audio sampled at 44.1 kHz with 16-bit precision in .wav format. 
%
\subsubsection{Background Noise Reduction}
The audio signals from in-home speech exercises are highly contaminated with time-varying background noise. Authors developed a method for reducing non-stationary noise in speech~\cite{cohen2003noise}. The audio signal is enhanced using noise estimates obtained from minima controlled recursive averaging. We performed a subjective evaluation for validating the suitability of this algorithm for our data. The enhanced speech was later used for extracting perceptual speech features such as loudness, fundamental frequency, jitter, frequency modulation, speech rate and sensory pleasantness (sharpness). We used the method developed in~\cite{cohen2003noise} for reducing non-stationary background noise in speech. It optimized the log-spectral amplitude of the speech using noise estimates from minima-controlled recursive averaging. Authors used two functions for accounting the probability of speech presence in various sub-bands. One of these functions was based on the time frequency distribution of apriori signal-to-noise ratio (SNR) and was used for estimation of the speech spectrum. The second function was decided by the ratio of the energy of noisy speech segment and its minimum value within that time window. Objective, as well as subjective evaluation, illustrated that this algorithm could preserve the weak speech segments contaminated with a high amount of noise~\cite{cohen2003noise}. Figure~\ref{fig_orig_enh_spec} shows the spectrogram of acquired speech signal from in-home trials and the spectrogram of corresponding enhanced speech signal. Speech enhancement is clearly visible in the darker regions (corresponding to speech) and noise reduction in lighter regions (corresponding to silences/pauses).
\begin{figure}[!t]
\centering
\includegraphics[width=350pt]{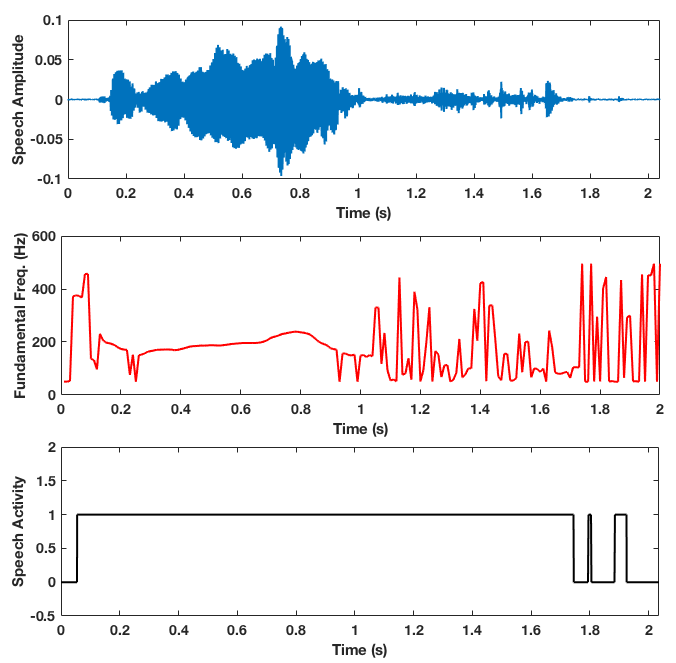}
\caption{Top sub-figure shows time-domain enhanced speech signal. The middle sub-figure depicts corresponding fundamental frequency contour. The bottom sub-figure shows the speech activity labels where '1' stands for speech and '0' for silence/pauses. We used speech activity detection proposed in ~\cite{tan2010low}. This is effective and has low computational expense.}
\label{fig_time_pitch_vad}
\end{figure}
\subsubsection{Automated Trimming of the Speech Signal}
We used the method developed in~\cite{tan2010low} for automated trimming of audio files by removing the non-speech segments. This method was validated to be accurate even at low SNRs that is typical for in-home audio data. The low computational complexity of this algorithm qualifies it for implementation on Fog device with limited resources. After applying the noise reduction method on acquired speech signal, we used voice activity detection (VAD) algorithm for removing the silences. Authors in~\cite{tan2010low} proposed a simple technique for VAD based on an effective selection of speech frames. The short time-windows of a speech signal are stationary (for 25-40 ms windows). However, for an extended time duration (more than 40 ms), the statistics of speech signal changes significantly rendering unequal relevance of speech frames. It necessitates the selection of effective frames on the basis of posteriori signal-to-noise ratio (SNR). The authors used energy distance as a substitute to the standard cepstral distance for measuring the relevance of speech frames. It resulted in reduced computational complexity of this algorithm. Figure~\ref{fig_time_pitch_vad} illustrates automated trimming of a speech signal for removing the pauses present in the audio files. We used time-windows of size 25 ms with 10 ms skip-rate between successive windows. 
%
\subsubsection{Fundamental Frequency Estimation}
We used the method proposed in~\cite{gonzalez2014pefac} for estimation of the fundamental frequency. It was found to be effective even at very low SNRs. It is a frequency-domain method referred as Pitch Estimation Filter with Amplitude Compression (PEFAC). We used 25 ms time-windows with 10 ms skip-rate for estimation of the fundamental frequency. In the first step, noise components were suppressed by compressing the speech amplitude. In the second step, the speech was filtered such that the energy of harmonics was summed. It involved filtering of power spectral density (PSD) followed by picking the peaks for estimation of the fundamental frequency (in Hz). Figure~\ref{fig_time_pitch_vad} shows the time-domain speech signal along with automatic trimming decision and pitch estimates for each overlapping windows. 

Another method we implement for fundamental frequency estimation is based on harmonic models~\cite{deb5_asgari2012robust}. Voiced speech is not just periodic but also rich in harmonic, so voiced segments are modeled by adopting harmonic  models. 
\begin{figure}[!t]
\centering
\includegraphics[width=350pt]{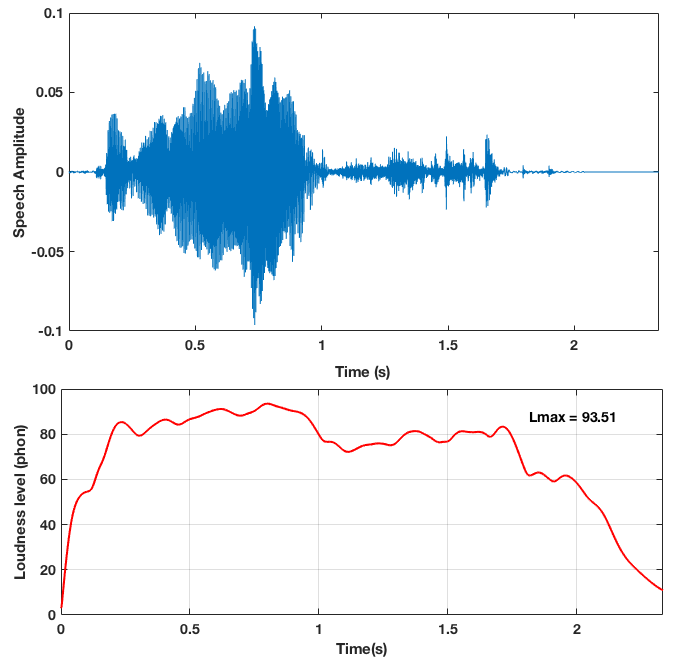}
\caption{The time-domain speech signal and corresponding instantaneous loudness curve. Loudness was computed over short windows of 25ms with 10 ms skip-rate.}
\label{fig_time_loud}
\end{figure}
\subsubsection{Perceptual Loudness}
Speech-language pathologists (SLPs) use loudness as an important speech feature for quantifying the perceptual quality of clinical speech. It is a mathematical quantity computed using various models of the human auditory system. There are different models available for loudness computation valid for specific sound types. We used Zwicker model for loudness computation valid for time-varying signals~\cite{zwicker2013psychoacoustics}. The loudness is perceived intensity of a sound. The human ears are more sensitive to some frequencies than the other. This frequency selectivity is quantified by the Bark-scale. The Bark scale defines the critical bands that play an important role in intensity sensation by the human's ears. The specific loudness of a frequency band is denoted as $L_{0}$ and measured in units of Phon/Bark. The loudness, L, (in unit Phon) is computed by integrating the specific loudness, $L_{0}$, over all the critical-band rates (on bark scale). Mathematically, we have
\begin{equation}
L = \sum_{0}^{24 Bark} L_{0} \cdot dz
\end{equation}
Typically, the step-size, dz, is fixed at 0.1~\cite{zwicker2013psychoacoustics}. We used Phon (in dB) as the unit of loudness level. Figure~\ref{fig_time_loud} shows a time-domain speech signal and corresponding instantaneous loudness in dB Phon. It depicts the dependence of loudness on speech amplitude.
\begin{figure}[!t]
\centering
\includegraphics[width=250pt]{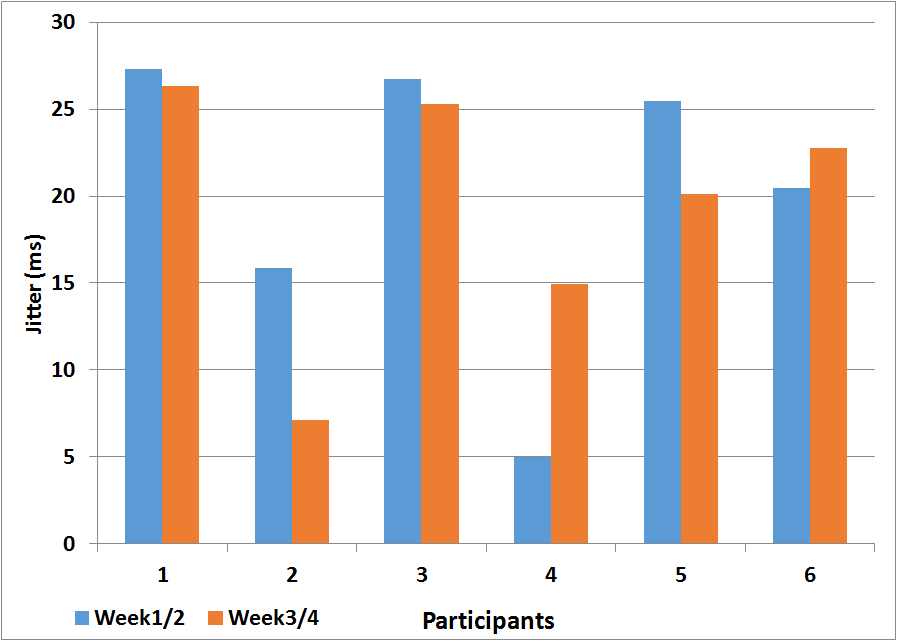}
\caption{The average jitter, $J_{1} $ (ms) computed using speech samples from six patients with Parkinson's disease who participated in field-trial that lasted four weeks. Three patients used Fog for first and third week while other three patients used it for second and fourth week. We are comparing weeks where Fog was used.}
\label{fig_jitter}
\end{figure}
\subsubsection{Jitter}
Jitter ($J_{1} $) quantifies changes in the vocal period from one cycle to another. Instantaneous Fundamental frequency was used for computing the jitter ~\cite{tsanas2012accurate}. $J_{1} $ was defined as the average absolute difference between consecutive time-periods. Mathematically, it is given as:
\begin{equation}
J_{1}= \frac{1}{M}\sum_{j=1}^{M-1} \vert F_{j} - F_{j+1} \vert
\end{equation}
where $F_{j} $ was the j-th extracted vocal period and $M$ is the number of extracted vocal periods.

Figure~\ref{fig_jitter} shows the comparison of jitter of six patients with PD from home-trials. Three patients used the Fog for first week and third week of the trial-month. Another three patients used Fog for second and fourth week. This swapping was done to see the effect of Fog architecture. In absence of Fog device, data was stored in android tablet (gateway) device and later was processed in offline mode. In presence of Fog, the data was processed online. Since same program produced these results, we can compare them. Figure~\ref{fig_jitter} shows the Jitter (in ms) for all cases. We can see that the change in jitter from first/second to third/fourth week is complicated. In some cases it increases while in other it decreases. Only specialized clinicians can interpret such variations. The Fog architecture facilitate the computation of jitter and sync it to cloud backend. Speech-language pathologists (SLPs) can later access these charts and correlated it with corresponding patient's treatment regime.
\subsubsection{Frequency Modulation}
It quantifies the presence of sub-harmonics in speech signal. Usually, speech signals with many sub-harmonics lead to a more complicated interplay between various harmonic components making it relevant for perceptual analysis. Mathematically, it is given as~\cite{sun2000pitch}:
\begin{equation}
F_{mod}= \frac{ \max \left( F_{j}\right)_{j=1}^{M} - \min \left( F_{j}\right)_{j=1}^{M}}{\max \left( F_{j}\right)_{j=1}^{M} + \min \left( F_{j}\right)_{j=1}^{M}
}
\end{equation}
where $F_{mod}$ is frequency modulation, and $ F_{j} $ is the fundamental frequency of j-th speech frame. 
\subsubsection{Frequency Range}
The range of frequencies is an important feature of speech signal that quantifies its quality~\cite{banse1996acoustic}. We computed the frequency range as the difference between $5-th$ and $95-th$ percentiles. Mathematically, it becomes: 
\begin{equation}
F_{range}= F_{95\%} - F_{5\%} 
\end{equation}
Taking $5-th$ and $95-th$ percentiles helps in eliminating the influence of outliers in estimates of fundamental frequency that could be caused by impulsive noise and other interfering sounds. 
\subsubsection{Harmonics to Noise Ratio}
Harmonics to Noise Ratio (HNR) quantifies the noise present in the speech signal that results from incomplete closure of the vocal folds during speech production process~\cite{tsanas2012accurate}. We used method proposed in~\cite{boersma1993accurate} for HNR estimation. The average and standard deviation of the segmental HNR values are useful for perceptual analysis by speech-language pathologist. Lets assume that $R_{xx}$  is normalized autocorrelation and $ l_{\max} $ is the lag (in samples) at which it is maximum, except the zero lag. Then, HNR is mathematically given by~\cite{boersma1993accurate}:   
\begin{equation}
HNR_{dB}=  10 \log10 \left( \frac{ R_{xx}(l_{\max})}{ 1- R_{xx}(l_{\max})} \right)
\end{equation}
%
%
\subsubsection{Spectral Centroid} 
It is the \textit{center of mass} of spectrum. It measure the \textit{brightness} of an audio signal.  Spectral centroid of a spectrum-segment is given by average values of frequency weighted by amplitudes, divided by the sum of amplitudes~\cite{paliwal1998spectral}. Mathematically, we have  
\begin{equation}
SC=\frac{\sum _{n=1}^{N} k F[k]}{\sum _{n=1}^{N} F[k]}
\end{equation}
where SC is the spectral centroid, and $ F[k] $ is amplitude of $k-th$ frequency bin of discrete Fourier transform of speech signal.
%
\subsubsection{Spectral Flux}
It quantifies the rate of change in power spectrum of speech signal. It is calculated by comparing the normalized power spectrum of a speech-frame with that of other frames. It determines the timbre of speech signal~\cite{yang2008regression}.
\subsubsection{Spectral Entropy}
We adopted it for speech-language pathology in this chapter. It is given by:
\begin{equation}
SE=\frac{-\sum P_{j} log(P_{j})}{log(M)}
\end{equation}
where SE is the spectral entropy, $ P_{j} $ is the power of j-th frequency-bin and M is the number of frequency-bins. Here, $\sum P_{k} = 1 $ as the spectrum is normalized before computing the spectral entropy. 
\subsubsection{Spectral Flatness} 
It measures the flatness of speech power spectrum. It quantifies how similar the spectrum is to that of a noise-like signal or a tonal signal. Spectral Flatness (SF) of white noise is 1 as it has constant power spectral density (PSD). A pure sinusoidal tone has SF close to zero showing the high concentration of power at a fixed frequency. Mathematically, SF is ratio of geometric mean of power spectrum to its average value~\cite{johnston1988transform}.
\begin{figure}[!t]
\centering
\includegraphics[width=250pt]{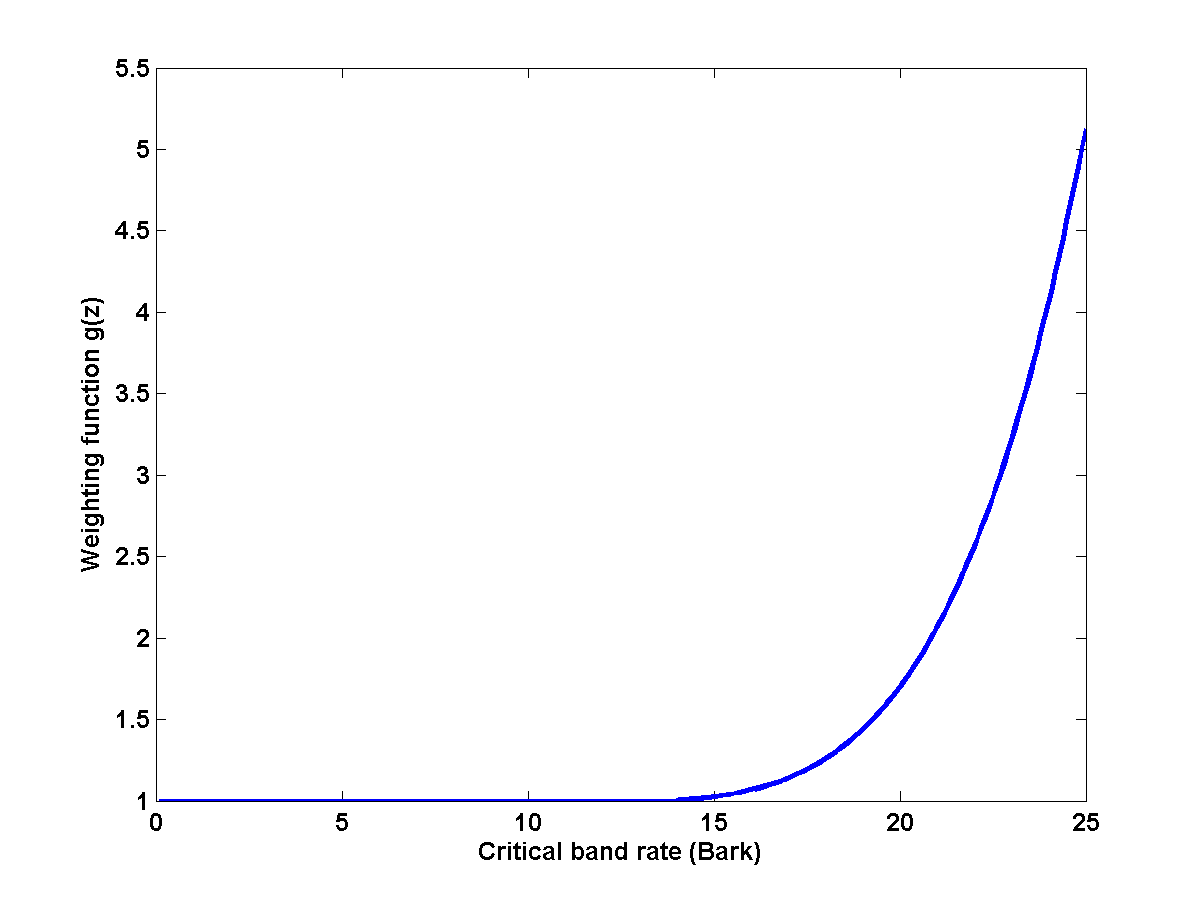}
\caption{The weights that were used for computing sharpness based on~\cite{zwicker2013psychoacoustics}. Sharpness quantifies perceptual pleasantness of the speech signal. We can see that the higher critical band rates use lower weights for computing the sharpness.}
\label{fig_weight_sharpness}
\end{figure}
\subsubsection{Sharpness}
Sharpness is a mathematical function that quantifies the sensory pleasantness of the speech signal. High sharpness implies low pleasantness. It value depends on the spectral envelope of the signal, amplitude level and its bandwidth. The unit of sharpness is acum (Latin expression). The reference sound producing 1 acum is a narrowband noise, one critical band wide with 1 kHz center frequency at 60 dB intensity level~\cite{fastl2007psychoacoustics}. Sharpness, $S$ is mathematically defined as
\begin{equation}
S= 0.11 \frac{\sum_{0}^{24 Bark}L_{0} \cdot g(z) \cdot z \cdot dz}{\sum_{0}^{24 Bark}L_{0}\cdot dz} acum
\label{eqn_sharpness}
\end{equation}
However, its  numerator is weighted average of specific loudness ($L_{0}$)
over the critical band rates. The weighting function, $g(z)$, depends on critical band rates. The $g(z)$ could be interpreted as the mathematical model for the sensation of sharpness shown in Figure~\ref{fig_weight_sharpness}. 
\begin{figure}[!tb]
\centering
\includegraphics[width=250bp]{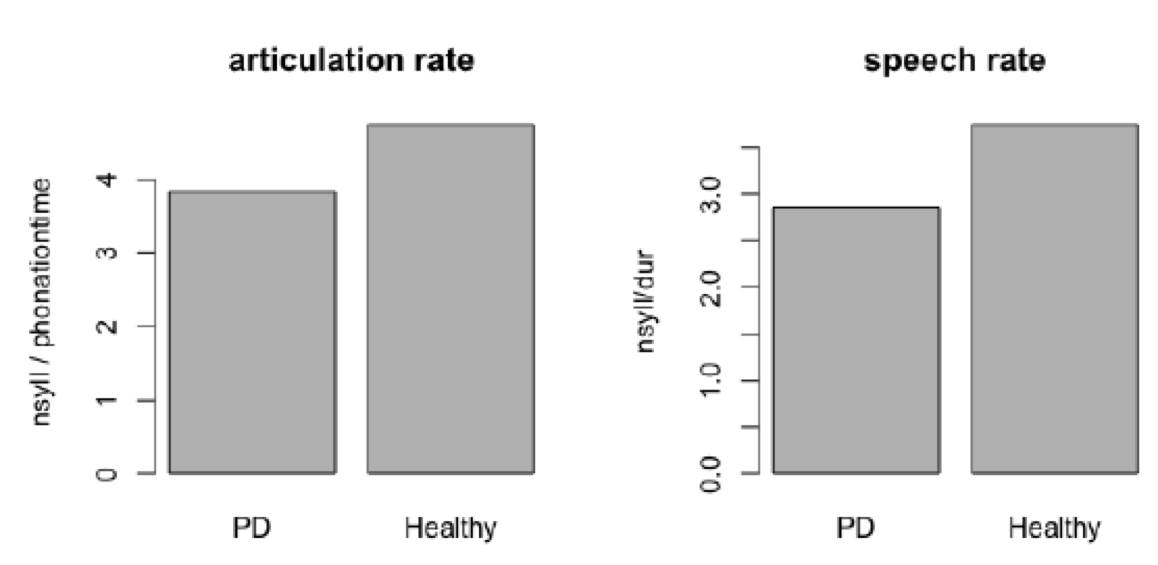}
\caption{The left sub-figure (a) shows the articulation rate (nsyll/phonation time) for the patients with PD and healthy controls. It shows that the healthy controls exhibit significantly higher articulation rate as compared to the patients with PD that is in accordance with the findings in~\cite{deb6_martinez2015speech}. The right sub-figure depicts speech rate for the same case. The y-axis represents the speech rate (number of syllables/duration) for the healthy controls and the patients with PD. The findings were that, Healthy control showed a higher speech rate as compared to the patients with Parkinson's disease. Speech rate for the healthy control was, 3.74 and for the PD subject 2.86. The analysis is done using Praat~\cite{praat} and the bar graph plots were  generated using R statistical analysis software.}
\label{fig_speech_rate}
\end{figure}
\begin{figure*}[!tb]
\centering
\includegraphics[width=480pt]{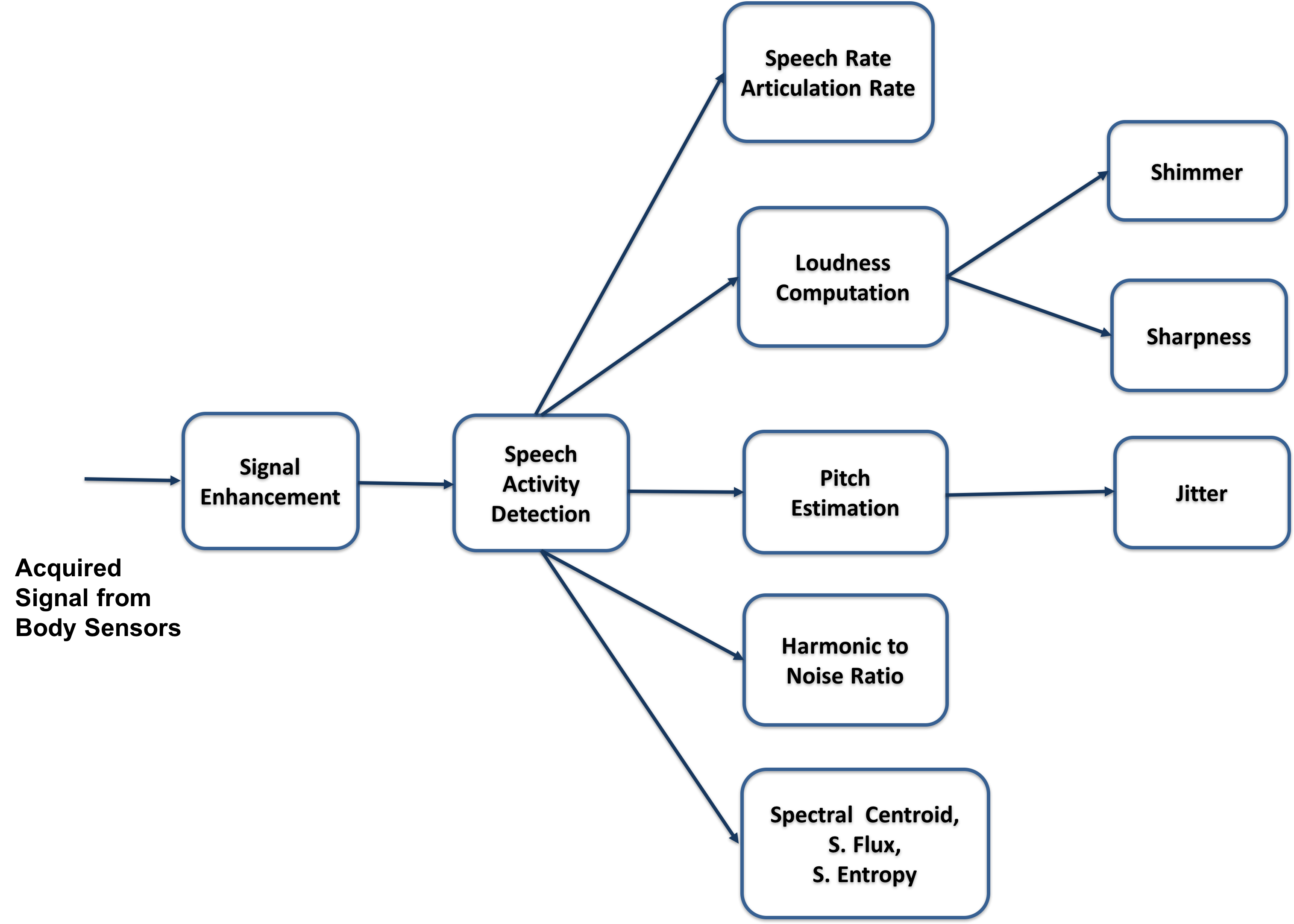}
\caption{Block diagram of pathological speech processing module in proposed Fog architecture. The speech signal is first enhanced to reduce the non-stationary background noise. Next, speech activity is detected to identify the speech regions and discard pauses/silences. Speech activity detection reduces the computation by ignoring non-speech frames. Finally, the speech is used for computing clinically relevant features using mathematical models of auditory perception.}
\label{fig_speech_proc}
\end{figure*}
\subsubsection{Speech Rate and Articulation Rate}
Praat scripting is extensively used in speech analysis. Some analysis were done using Praat scripting language. Slurred speech, breathy and hoarse speech, difficulty in fast-paced conversations are some of the symptoms of Parkinson's disease. The progressive decrease in vocal sonority and intensity at the end of the phonation is also observed in patients with PD~\cite{deb6_martinez2015speech}. Literature suggests that speech and articulation rates decrease in PD, and there is a causal link between duration and severity of PD with this decrease in articulation rate~\cite{deb6_martinez2015speech}. Articulation rate is a prosodic feature and is defined as a measure of rate of speaking excluding the pauses. Speech rate is usually defined as the number of sounds a person can produce in a unit of time~\cite{deb6_martinez2015speech}. As illustrated in~\cite{praat}, Speech rate is calculated by detecting syllable nuclei. We used Wempe's algorithm for estimating the speech rate~\cite{praat}. For analysis of speech rate and articulation rate, Praat scripts were used. Two sound samples were chosen for comparative analysis. Samples comprised of healthy control and the patients with PD. Figure~\ref{fig_speech_rate} shows the bar-chart for articulation rate and speech rate.

\begin{figure}[!t]
\centering
\includegraphics[width=350bp]{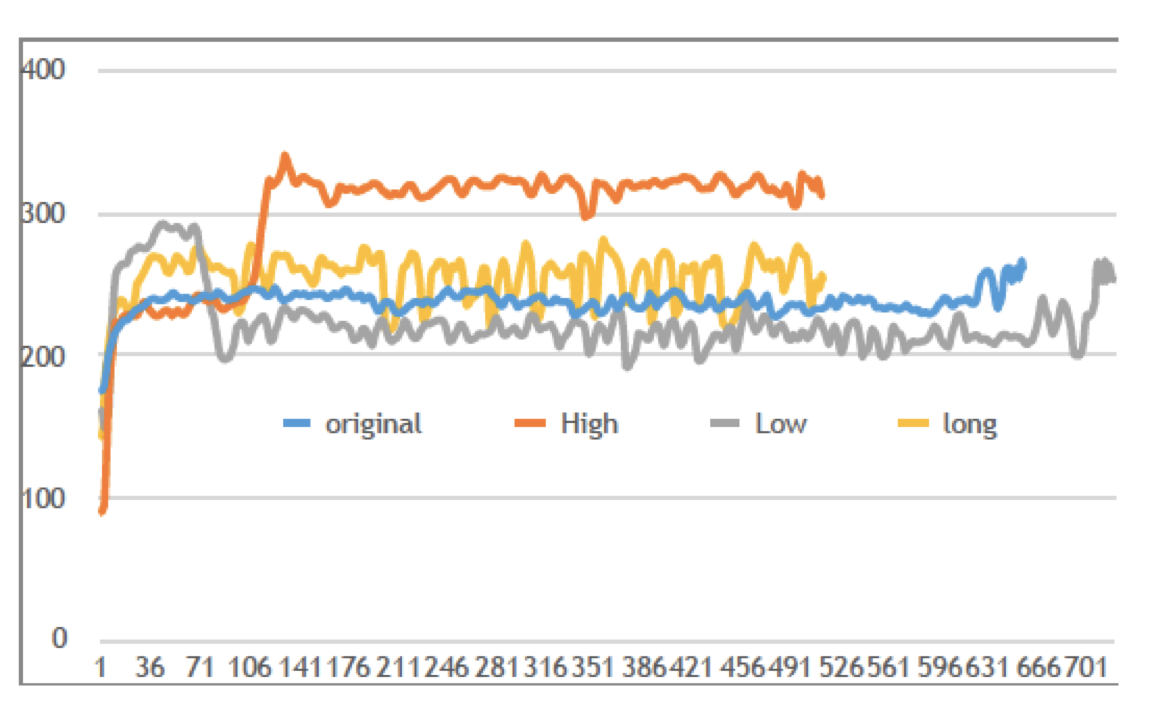}
\caption{Depicting the variations in  frequency in sustained /a/ (task $t_{1}$), HIGHS (task $t_{2}$), and LOWS (task $t_{3}$) for several speech samples.}
\label{fig_deb2}
\end{figure}
\begin{figure}[!t]
\centering
\includegraphics[width=350pt]{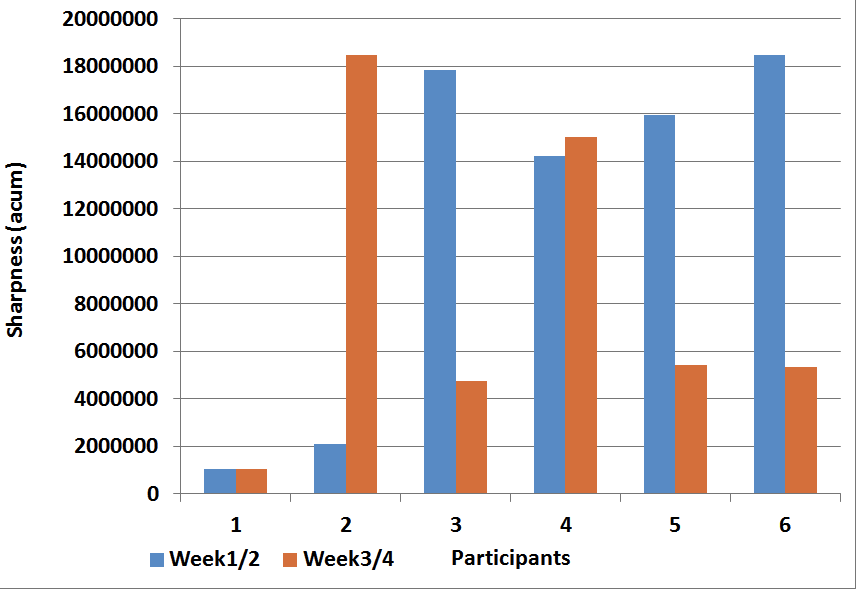}
\caption{The comparison of average sharpness of the
speech signal obtained from in-home trails of a patient. The six
days of two weeks are compared with respect to average
sharpness (in acum). These two weeks are separated by one
week. Low sharpness shows high sensory pleasantness in a
speech signal. We can see that the evolution of sharpness on
different days is very complicated even during the same week. It
is because the speech disorders are unique for each patient with PD.}
\label{fig_weight_sharpness}
\end{figure}
\begin{figure*}[!tb]
\centering
\includegraphics[width=490pt]{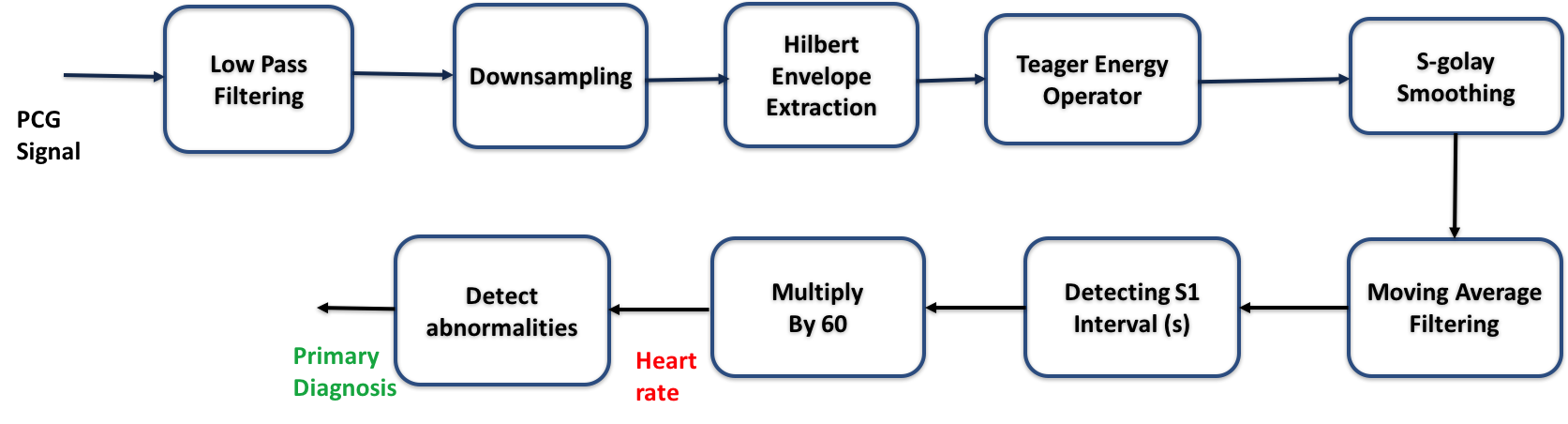}
\caption{Proposed method for estimation of heart rate from PCG signal. We first do low pass filtering for reducing the high frequency noise. It is followed by downsampling for reducing the computational complexity. Next, Hilbert envelope is extracted and envelope is processed with Teager energy operator (TEO). The output of TEO is smoothed by Savitzky-Golay filtering. We performed moving averaging for further enhancement of peaks corresponding to heart sound S1. The time-period of heart sound S1 (in seconds) is multiplied with 60 get the heart rate in Beats per minute (BPM). The normal heart rate lies in range 70-200 BPM. Significant deviation from this range shows abnormality in cardiac cycle. This method was implemented in Python and executed in the Fog computer. }
\label{fig_pcg_proc}
\end{figure*}
\begin{figure*}[!tb]
\centering
\includegraphics[width=490pt]{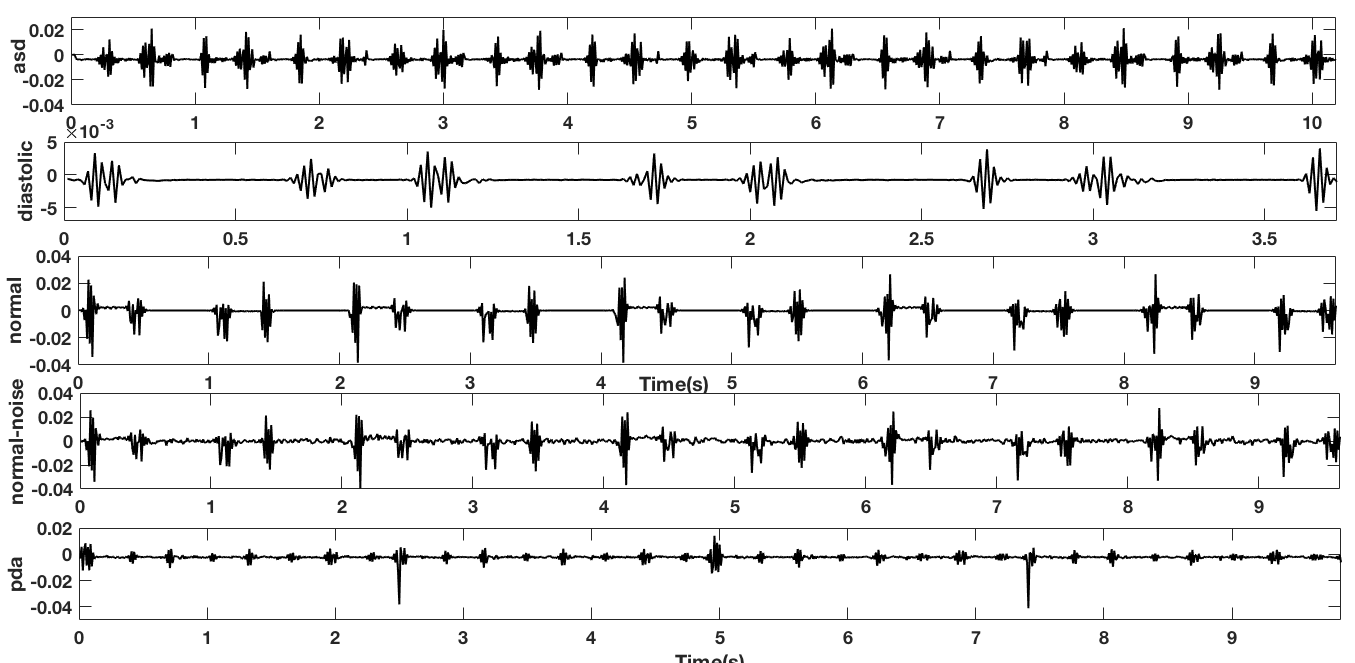}
\caption{Time-domain PCG signal for four conditions namely normal, asd, pda, diastolic. The variations in these signals reflect the corresponding cardiac functions.}
\label{fig_pcg_time}
\end{figure*}
\begin{figure*}[!tb]
\centering
\includegraphics[width=490pt]{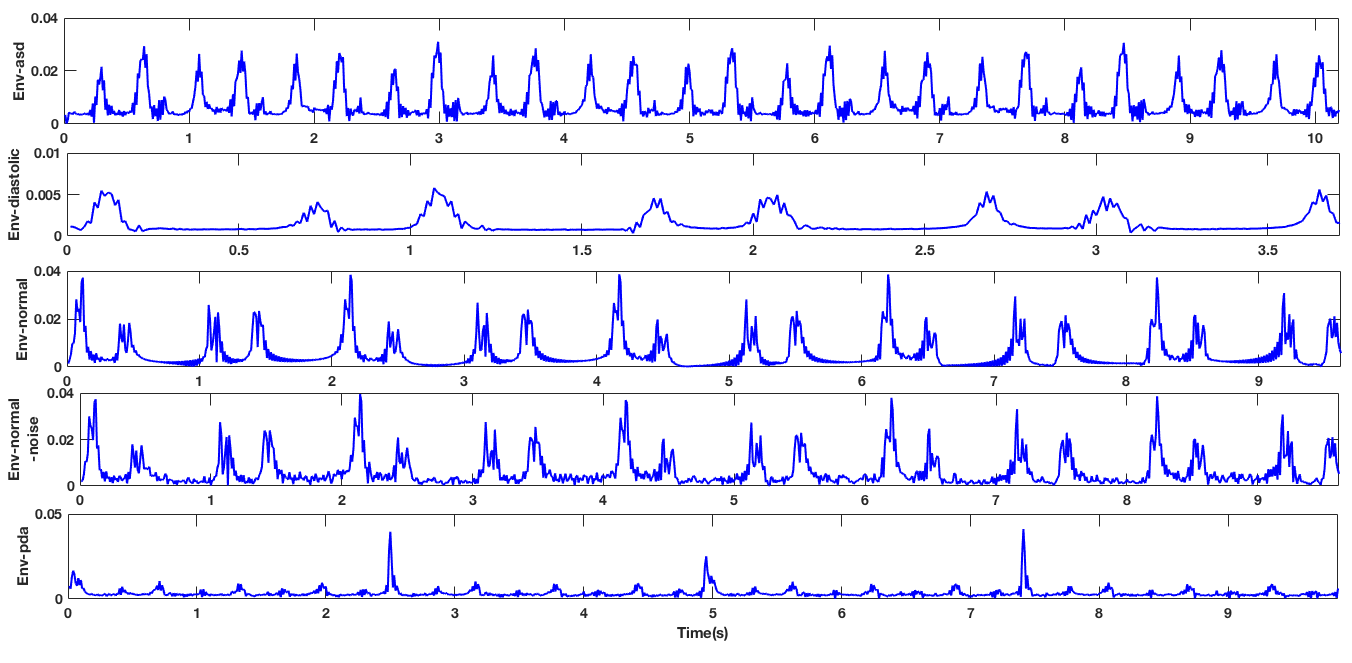}
\caption{Envelope of the PCG signal using procedure shown in block diagram (see Figure~\ref{fig_pcg_proc}) for four conditions namely normal, asd, pda, diastolic. The envelope shows clear transitions in PCG signal that can be further processed for localizing the fundamental sound S1 and hence estimation of heart rate in Beats per minute (BPM).}
\label{fig_pcg_env}
\end{figure*}
\subsection{Case Study II: Phonocardiography (PCG)-based Heart Rate Monitoring}
Phonocardiography refers to acquisition of heart sounds that contains signatures of abnormalities in cardiac cycle. There are two major sound, S1 and S2 associated with cycle of cardiac rhythm. Traditionally, specialized clinicians listen heart sound using devices such as stethoscopes for cardiac diagnosis. Such examination need specialized training~\cite{geddes2005birth}. Authors developed a computationally inexpensive method for preliminary diagnosis of heart sound~\cite{reed2004heart}. Segmentation of PCG signals and estimation of heart rate from it has been done primarily using two approaches. Segmentation of PCG signals and estimation of heart rate from it has been done primarily using two approaches. The first approach uses ECG as a reference for synchronization of cardiac cycles. Second approach relies solely on PCG signal and is appropriated for wearable devices that relies on smaller number of sensors.
 
In this paper, we integrated the analysis method into Fog framework for providing local computing on Fog device. With the growing use of wearables~\cite{brusco2005development} for acquiring PCG data, there is need of processing such data for preliminary diagnosis. Such preliminary diagnosis refers to segmentation of PCG signal into heart sounds S1 and S2 and extraction of heart rate. Figure~\ref{fig_pcg_proc} shows the proposed scheme for analysis of PCG data for extracting the heart rate. We detect the time-points for heart sounds S1 and S2. Later, these were used for extracting the heart rate. The development and execution of a robust algorithm on Fog device is novel contribution of this chapter. 
%
\subsubsection{PCG Data Acquisition}
PCG signals were acquired using a wearable microphones kept closer to the chest. Such wearable devices could send data to a nearby placed fog device through a smartphone/tablet (gateway). Fog saves the PCG data in .wav format sampled at 800 Hz with 16 bit resolution. The microsoft wav format is lossless format and is widely used for healthcare sound data.  We are not discussing the hardware details as our primary goal is computing signal features on Fog device. The segmentation step (see block diagram in Figure~\ref{fig_pcg_proc}) separated the heart sounds S1 and S2 from the denoised PCG signal. The heart sounds S1 and S2 captures the acoustic cues from cardiac cycle. The peak-to-peak time-distance between two successive S1 sounds make one cardiac cycle. Thus, time-distance between two S1 sound determines the heart rate. 

We used the data from four scenarios of cardiac cycles namely, normal, asd, pda, and diastolic. The 'normal' refers to normal heartbeat from an healthy person. The 'asd' refers to PCG data induced by an atrial septal defect (a hole in the wall separating the atria). The 'pda' refers to PCG signal induced by patent ductus arteriosus (a condition wherein a duct between the aorta and pulmonary artery fails to close after birth). The last one, 'diastolic' refers to PCG signal corresponding to a diastolic murmur (leakage in the atrioventricular or semilunar valves). Figure~\ref{fig_pcg_time} shows the time domain PCG signals corresponding to these scenarios. We can see that PCG signal contain signatures of cardiac functioning and clear distinction is portrayed by these time-domain signals. Figure~\ref{fig_pcg_env} shows the enveloped of these signals (see Figure~\ref{fig_pcg_proc}). We can see that envelope shows the better track of time-domain variations. 
\subsubsection{Noise Reduction in PCG data}
The PCG signal was acquired at 800 Hz for capturing high fidelity data. Some noise is inherently present in data collected using wearable PCG sensors. We do low pass filtering using a sixth-order Butterworth filter with a cutoff frequency of 100 Hz. It reduces the noise leaving behind spectral components of cardiac cycle. We downsampled the low-pass filtered signal to reduce the computational complexity. 
\begin{figure}[!tb]
\centering
\includegraphics[width=350pt]{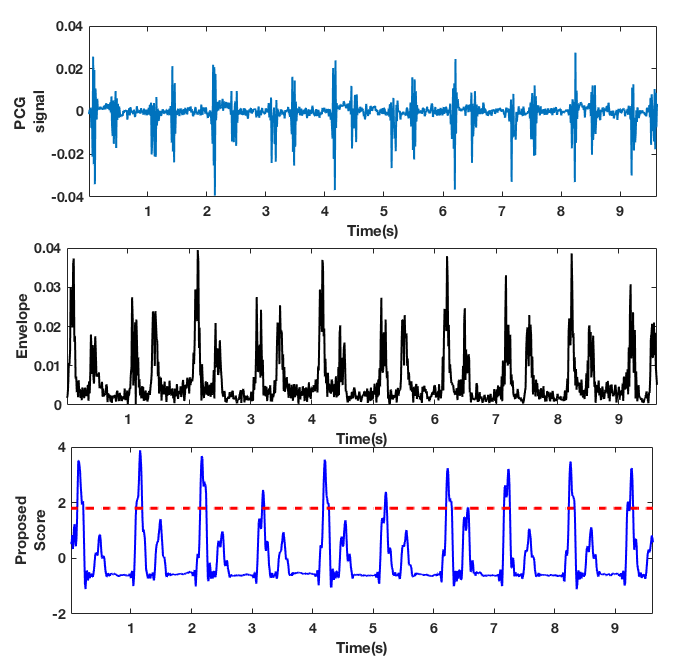}
\caption{Detecting heart sound S1 and using it for heart rate estimation in units of beats per minute (BPM). The top-figure shows the low pass and downsampled PCG signal. The middle figure shows the envelope by procedure shown in block diagram of Figure~\ref{fig_pcg_proc}. The bottom sub-figure shows the final post-processed envelope. It is clear that by choosing a suitable threshold, we can detect the S1 sound from PCG signal. Since the cardiac cycle time-length (in seconds) is same as time-difference between two S1 sound (in seconds), we can estimate heart rate by multiply it with 60.}
\label{fig_pcg_detect}
\end{figure}
\subsubsection{Teager Energy Operator for Envelope Extraction}
Teager Energy Operator (TEO) is a nonlinear energy function~\cite{kaiser1993some}. TEO captures the signal energy based on physical and mechanical aspects of signal production. It has been successfully in various applications~\cite{li2016improved,kvedalen2003signal}. For a discrete signal $x[n]$, it is given by
\begin{equation}
\Psi (x[n]) = x[n]*x[n] - x[n+1]*x[n-1]
\label{eqn_teo}
\end{equation}
where $\Psi (x[n])$ is the TEO corresponding to the sample $x[n]$. We applied TEO on the downsampled signal (see Figure~\ref{fig_pcg_proc}) to extract the envelope. The TEO output is further smoothed using Savitzky-Golay filtering. Savitzky-Golay filters are polynomial filters that achieve least-squares smoothing. These filters performed better than standard finite impulse response (FIR) smoothing filters.~\cite{orfanidis1995introduction}. We used fifth-order Savitzky-Golay smoothing filters with a frame-length of 11 windows. Next, we perform moving-average filtering on smoothed TEO envelope. The window-length of 11 was used for moving averaging. In next step, the output of moving-average filter is mean and variance normalized to suppress the channel variations. 
\subsubsection{Heart Rate Estimation by Segmentation of Heart Sounds S1}
The heart sound S1 marks the start of the systole. It is generated by closure of mitral and tricuspid valves that cause blood flow from atria to ventricle. It happens when blood has returned from the body and lungs. The heart sound S2 marks the the end of systole and the beginning of diastole. It is generated upon closure of aortic and pulmonary valves following which the blood moves from heart to the body and lungs. Under still conditions, the average heart-sound duration are S1 (70-150ms) and S2 (60-120ms). The cardiac cycle lasts for 800 ms where systolic period is around 300 ms and diastolic period being 500 ms~\cite{varghees2014novel}.

The mean and variance normalized envelope is used for detecting the fundamental heart sound (S1). Since S1 marks the span of cardiac cycle, we compute time-distance between two S1 locations. It gives the length of cardiac cycle (in seconds). This is multiplied by 60 (see Figure~\ref{fig_pcg_proc}) to get the heart rate in Beats per minute (BPM). Under normal cardiac functioning heart rate lies in range 70-200 BPM. In case, where estimated heart rate is significantly large than this range over a long duration of time, it shows some abnormality in health. It is worth to note that intense exercises such as running on treadmill, cycling~\emph{etc.} can also cause increase in heart rate. The Fog computer receives the PCG signals from wearable sensors and extract heart rate in BPM for each frame. We choose a time-windows of size two seconds with 70\% overlap between successive windows. 
\begin{figure}[!t]
 \centering
 \includegraphics[width=350pt]{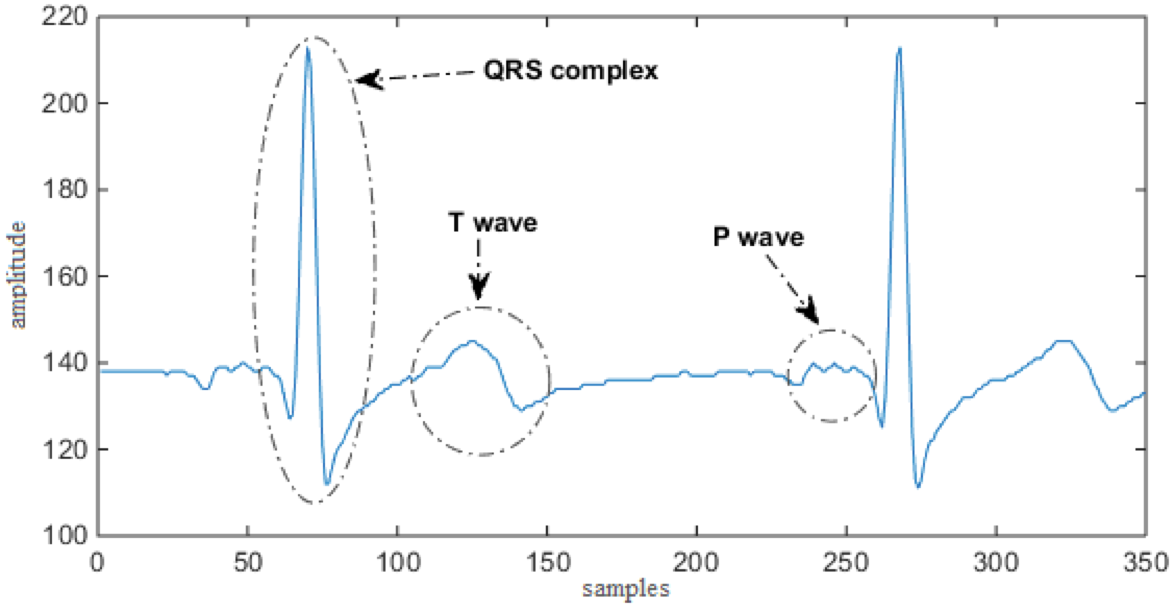}
 \caption{An example typical time-domain ECG waveform showing phases P, QRS complex and T.}
\label{fig_ecg_wav}
\end{figure}
\begin{figure}[!t]
 \centering
 \includegraphics[width=350pt]{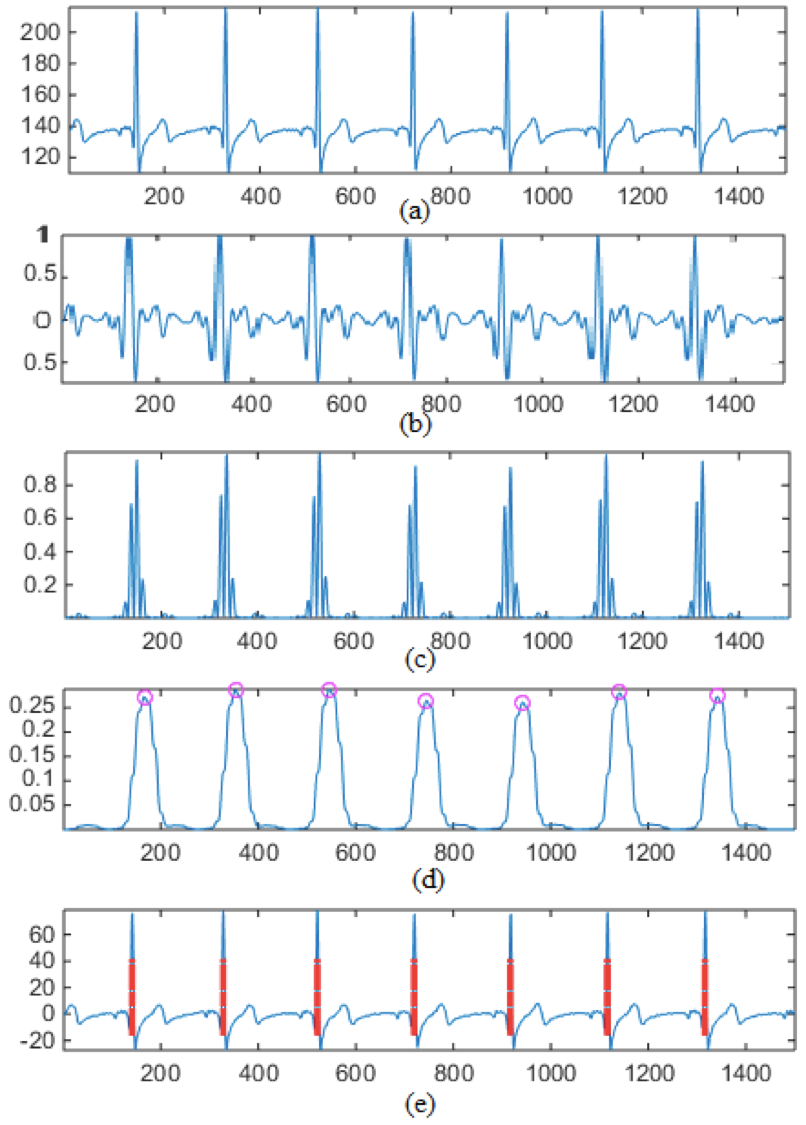}
 \caption{Illustration of QRS detection using Pan-Tompkins algorithm; (a) Raw
ECG data; (b) ECG signal after band-pass filtering and derivation; (c) Squaring the data; (d) Integration and thresholding to detect QRS; (e) Pulse train of ECG signal.}
\label{fig_pan_to}
\end{figure}
\subsection{Case Study III: Electrocardiogram (ECG) Monitoring}
Heart diseases are one of the major chronic illness with a dramatic impact on productivity of affected individuals and related healthcare expenses. An ECG sub-system is considerably for more out-of-hospital applications, manufacturers face continued pressure to reduce system cost and development time while maintaining or increasing performance levels. The electrocardiogram (ECG) is a diagnostic tool to assess the electrical and muscular functions of the heart. The ECG signal consists of components such as P wave, PR interval, RR interval, QRS complex, pulse train, ST segment, T wave, QT interval and infrequent presence of U wave. Presence of arrhythmias changes QRS complex, RR interval and pulse train. For instance a narrow QRS complex (<120 milliseconds) indicates rapid activation of the ventricles that in turn suggests that the arrhythmia originates above or within the his bundle (supraventricular tachycardia) and a wide QRS
(greater than 120 milliseconds) occurs when ventricular activation is abnormally slow. The most common reason for a wide QRS complex is arrhythmia of the ventricular myocardium (e.g., ventricular tachycardia)~\cite{alzand2011diagnostic}. Figure~\ref{fig_ecg_wav} shows ECG time series with P wave, T wave and QRS complex. These three patterns are search using DTW for a large number of ECG data sets. The last section of this case study will discuss the data reduction using DTW and GNU zip compression on ECG data.
The goal of our experiment is to detect arrhythmic ECG beats or QRS changes using QRS complex and the RR interval measurements. The ECG data is fed to the Fog computer from Internet-based database. The Fog computer extracts QRS
complex from ECG signals using real-time signal processing implemented in Python on Intel Edison. The Pan Tompkins algorithm is used for detection of QRS complex~\cite{pan1985real}. Pan-Tompkins algorithm consists of five steps:
\subsubsection{Band Pass Filtering}
The energy contained in QRS complex is approximated in 5-15 Hz range~\cite{alzand2011diagnostic}. We apply a band pass filter for extracting 5-
15 Hz content of ECG signals. The band pass filter reduces
muscle noise, 60 Hz power-line interference, baseline
wandering and T wave interference. This filter achieve a 3dB
pass-band from about 5-12 Hz. The high-pass filter is designed
by subtracting the output of first-order low-pass filter from an
all-pass filter with delay of 16 samples (80ms)~\cite{pan1985real}.
\subsubsection{Derivation}
The output of band-pass filter is differentiated to get the slope.
It uses a five-point derivative. After differentiation, the output signal is squared to get only positive values. It performs non-linear amplification of the
output suppressing the values lower than 1. A moving-window integration is applied on output of last step. It smoothens the output resulting in multiple peaks within duration of QRS complex. It adapts to changes in the ECG signal by estimating the signal and noise peaks for finding the R-peaks (Figure~\ref{fig_pan_to}). The Pan-Tompkins based QRS detection is implemented on ECG signals obtained from MIT-BIH Arrhythmia Database~\cite{mitdb}. Figure~\ref{fig_pan_to} illustrates the QRS detection using Pan-Tomkins algorithm on Intel Edison using MIT-BIH
Arrhythmia data. The ECG signal containing 2160 samples take 1 second of processing time on Intel Edison Fog computer. It shows that proposed Fog architecture is well suited for real-time ECG monitoring.
\begin{figure}[!t]
 \centering
 \includegraphics[width=350pt]{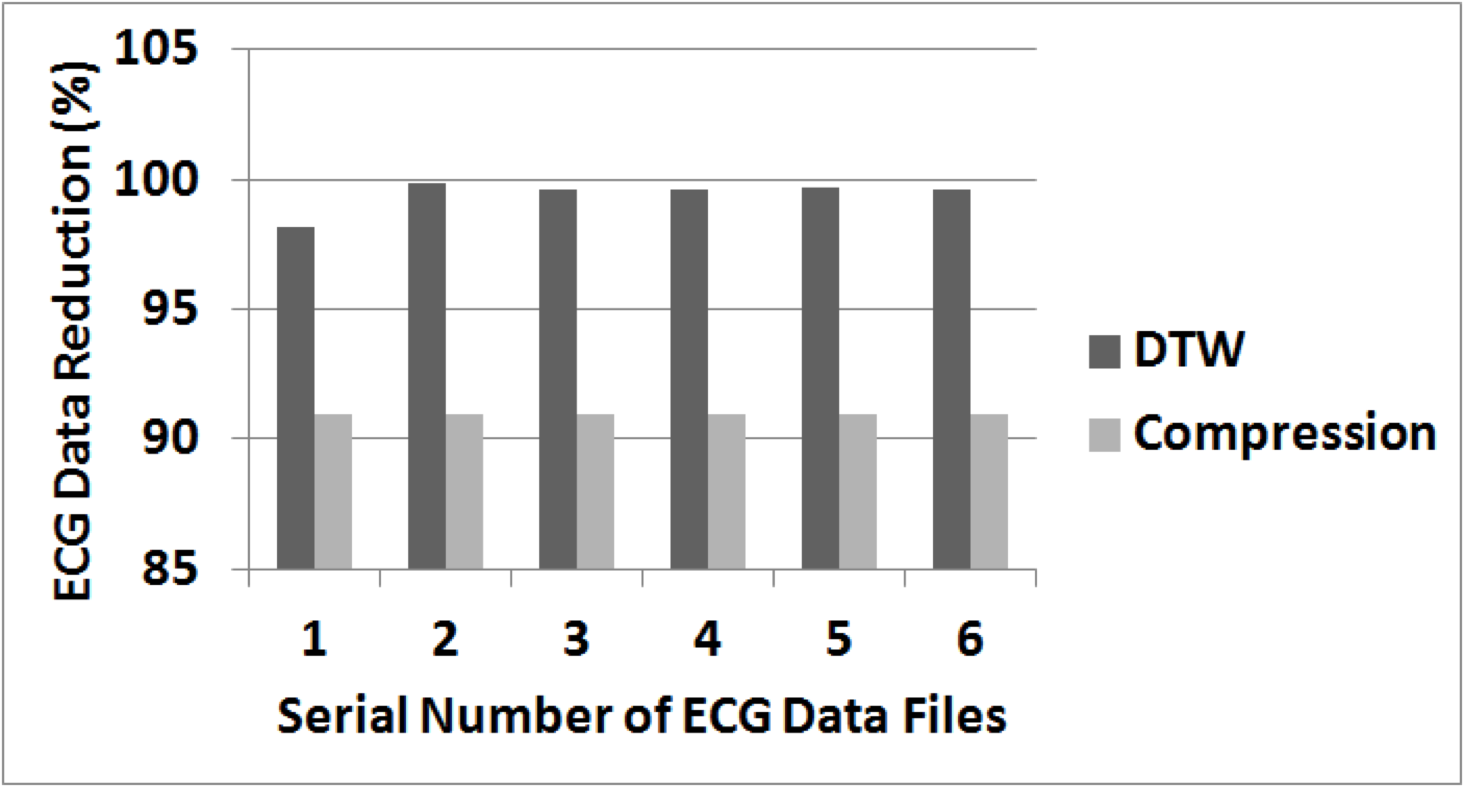}
 \caption{Comparison of data reduction resulting from DTW based pattern
mining and GNU zip based compression for ECG data obtained from MIT-BIH
Arrhythmia Database~\cite{mitdb}.}
\label{fig_ecg_data}
\end{figure}
\begin{figure}[!t]
 \centering
 \includegraphics[width=250pt]{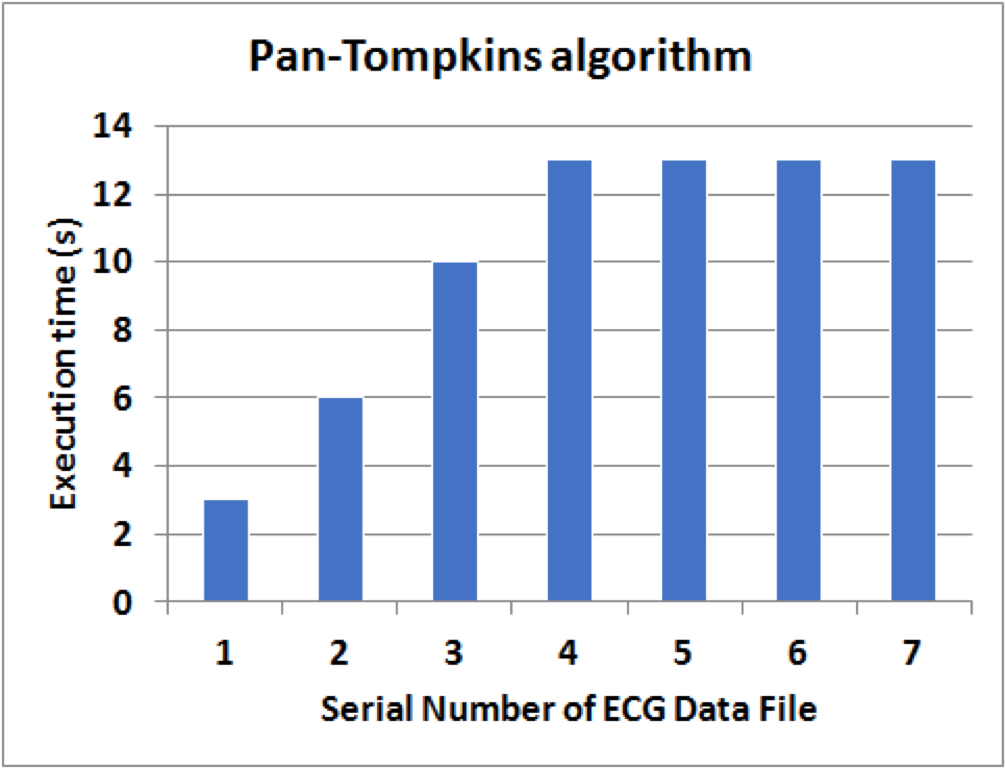}
 \caption{Execution time (in seconds) for Pan-Tompkins based QRS
detection on Inter Edison Fog computer for ECG data from MIT-BIH
Arrhythmia Database~\cite{mitdb}.}
\label{fig_ecg_time}
\end{figure}
We used DTW based pattern mining for P wave, T wave and QRS complex in ECG data. The DTW indices showing the location of these pattern in ECG time-series is sent to the cloud. Similarly, we use GNU zip program to compress the original ECG time series. The compressed ECG data files are then send to the cloud. Figure~\ref{fig_ecg_data} shows the data reduction resulting from DTW based pattern mining with compression. Similar to speech data, DTW reduces ECG data by more than 98\% in most of the cases while compression reduces around 91\%. Figure~\ref{fig_ecg_time} shows the execution time (in seconds) for Pan-Tompkins
based QRS detection implemented in Python on Intel Edison Fog computer. The data sets from MIT-BIH Arrhythmia Database are used. The size of the data sets range from 16.24 kB to 36.45 kB. The execution time increases with increase in file size. The time taken is always less than 15 seconds. This validates the efficacy of Fog Data architecture for real-time ECG monitoring.
\begin{figure*}[!t]
\centering
\includegraphics[width=480bp]{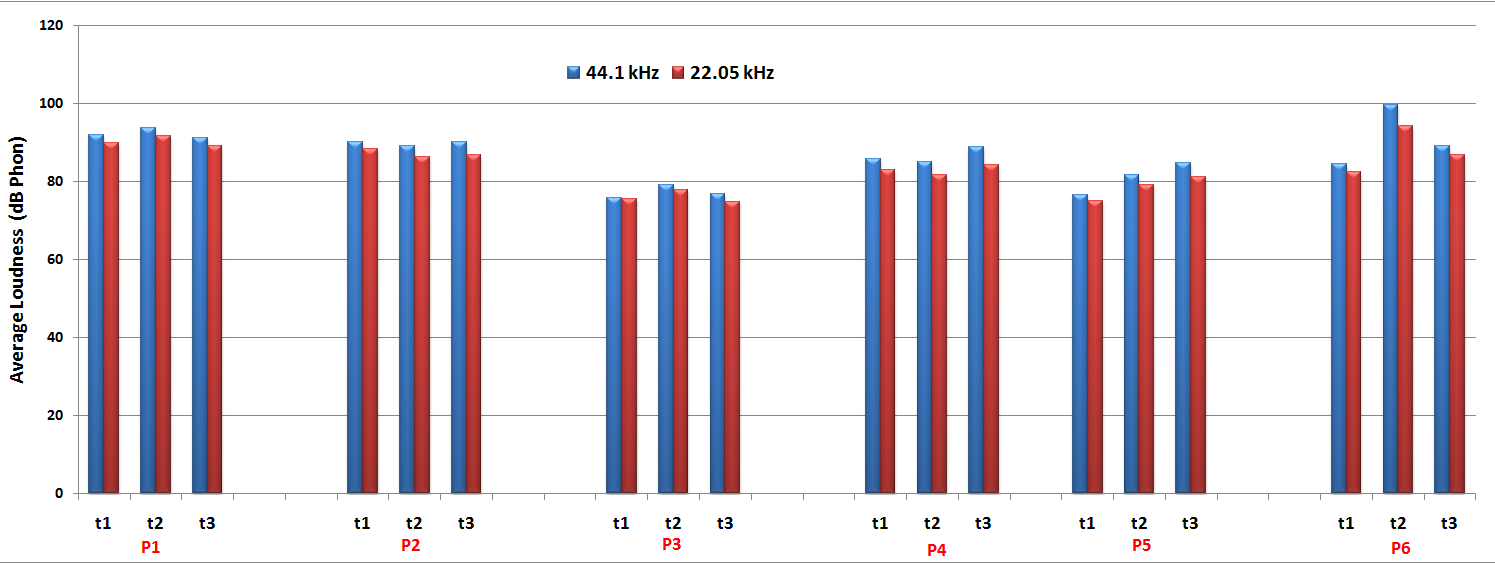}
\caption{Comparing loudness computed from speech signal recorded by smartwatch at sampling rate of 44.1 kHz and half of it. We can see the variations are low. The mean change with respect to 44.1 kHz is 2.86\% with a standard deviation of 1.26\%.} 
\label{fig_22k}
\end{figure*}
\begin{figure*}[!t]
\centering
\includegraphics[width=480bp]{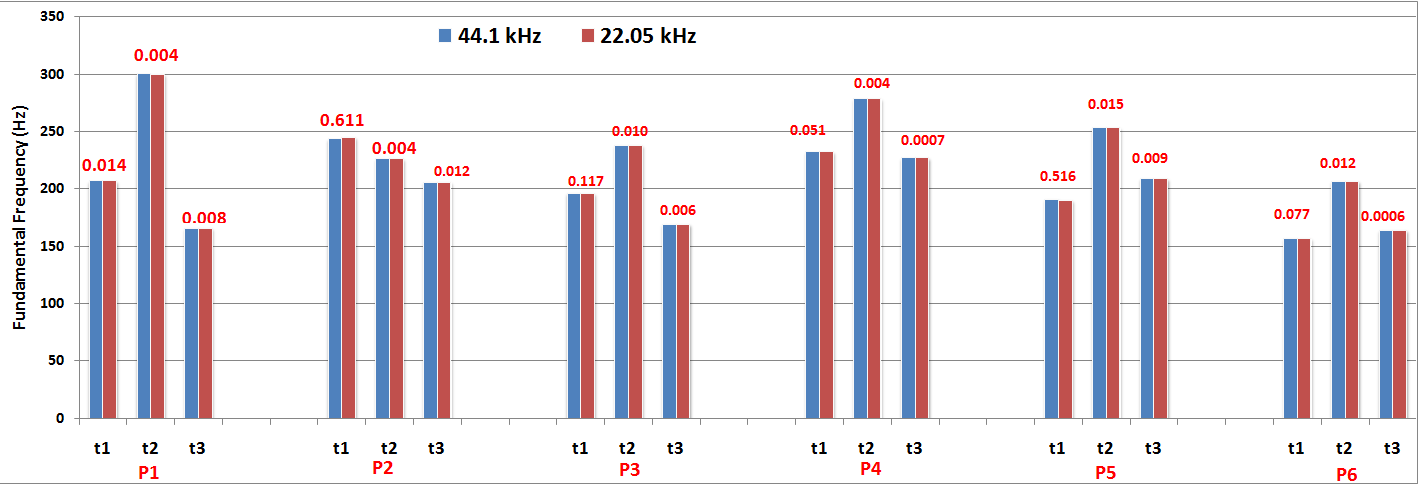}
\caption{Comparing fundamental frequency (in Hz) computed from speech signal recorded by smartwatch at sampling rate of 44.1 kHz and half of it. We can see the variations are low. The mean change with respect to 44.1 kHz is 0.0818
\% with a standard deviation of 0.1786 \%.} 
\label{fig_22k}
\end{figure*}
\begin{figure}[!t]
\centering
\includegraphics[width=350pt]{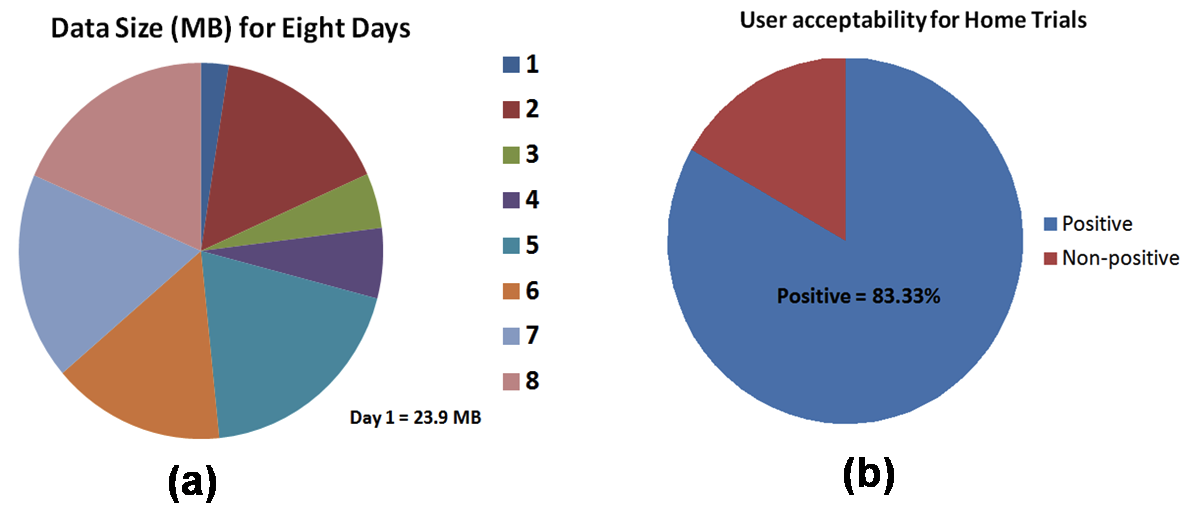}
\caption{(a) The data collected from one of the patients for 8 days. The speech data collected was well structured with date and time stamps.(b) The pie chart shows the user acceptability of the proposed system during in-home trials. By non-positive, we mean neither a positive nor a negative inclination towards
proposed system. For one participant with severe motor disorders, using smartwatch needed some effort and hence had neither a positive or negative inclination.}
\label{fig_data_stats}
\end{figure}
\begin{table}[!t]
\small
\centering
\caption{Latency measurements of Fog for computing the clinical speech features namely zero crossing rate (ZCR), special centroid (SC), and short-time energy (STE).}
\begin{tabular}{c c c c}
\hline
\textbf{Speech Tasks} &\textbf{Processing Time(s)} & \textbf{File Duration(s)} & \textbf{Size(kB)}\\
\hline
Task1 & 2.34 & 6.24 & 551\\
\hline
Task2 & 2.33 & 6.18 & 545\\
\hline
Task3 & 2.12 & 5.62 & 496\\
\hline
Task4 & 2.28 & 6.08 & 537\\
\hline
Task5 & 1.86 & 4.96 & 438\\
\hline
\textbf{Total} & 10.94 & 29.08 & 2567\\
\hline
\label{table3}
\end{tabular}
\end{table}
\begin{figure*}[!t]
\centering
\includegraphics[width=490pt]{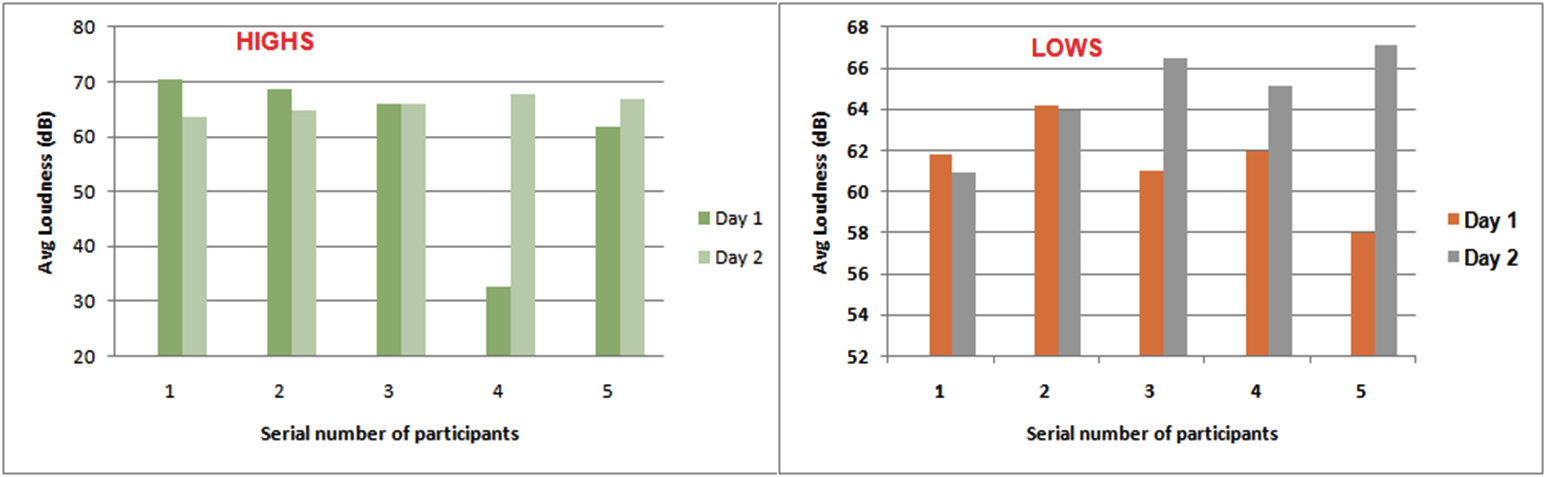}
\caption{The average loudness for two days on task "Highs" (task $t_{2}$)  and "Lows" (task $t_{3}$) for six patient doing speech exercises at home. The data was processed with Fog in real-time. It illustrates the Fog functionality to compute these features.}
\label{fig_high_low_loud}
\end{figure*}
\begin{figure*}[!t]
\centering
\includegraphics[width=490pt]{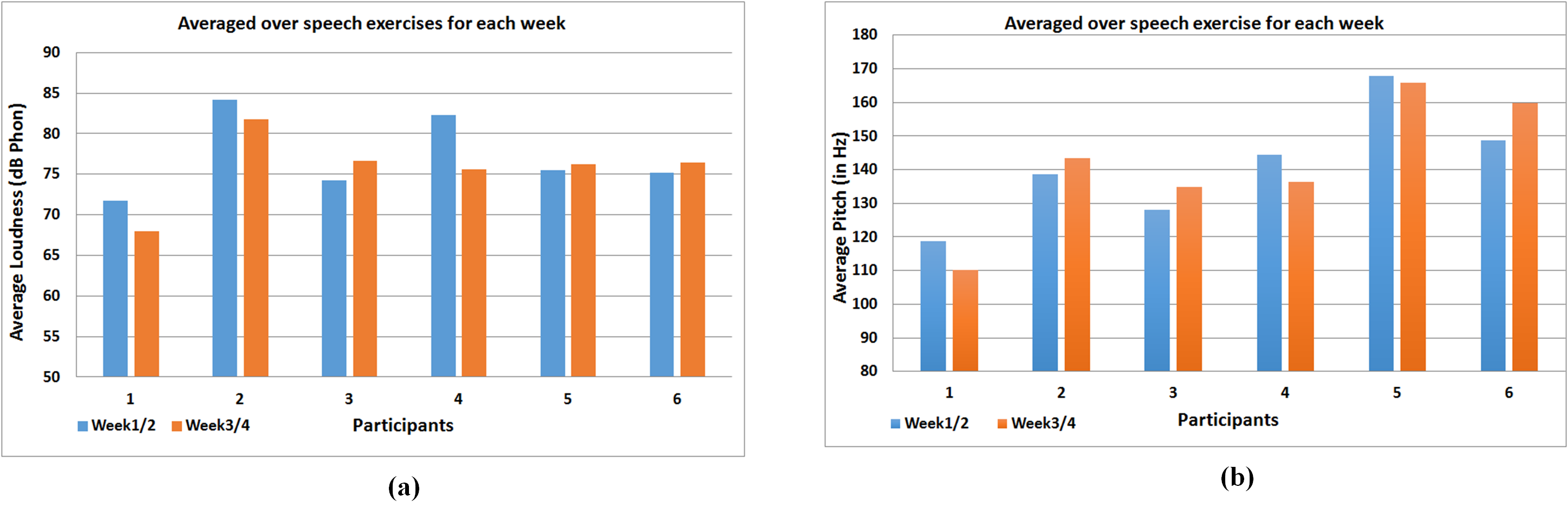}
\caption{Showing average loudness and pitch for each day for in-home trials for six patients. The patients used Fog for alternate weeks. We can see that each patient has a different trend for change in loudness and pitch. Interpretation of these variations is done by trained clinicians such as speech-language pathologists (SLPs). Fog compute these features and sync it to the secured cloud backend from where it can be accessed by SLPs, caregivers.}
\label{fig_home_loud_pitch}
\end{figure*}
\section{Experiments~\& Results}
\subsection{Intel Edison Description}
The Intel Edison platform used in this application was designed with a core system consisting of dual-core, dual-threaded Intel Atom CPU at 500MHz and a 32-bit Intel Quark microcontroller at 100MHz, along with connectivity interfaces capable of Bluetooth 4.0 and dual-band IEEE 802.11a/b/g/n via an on-board chip antenna. This platform came with a Linux environment called Yocto, which is not an embedded distribution of Linux itself, its true purpose is to provide an environment to develop a custom Linux distribution. We did not create a Linux distribution, instead we deployed a prebuilt distribution of Debian/Jessie for 32-bit systems. This decision was made such that we could deploy the same environment on both the Intel Edison and the Raspberry Pi.

\subsection{Raspberry Pi Description}
The Raspberry Pi Model B platform used in this application was designed with a core system consisting of a 900MHz 32-bit quad-core ARM Cortex-A7 CPU, and 1GB RAM. Since the Raspberry Pi does did not have WIFI connectivity built-in a WIFI dongle based on the Realtek RTL8188CUS chipset was installed. This platform came with a custom Linux distribution called Raspbian. Since Raspbian would provide a slightly different environment it was replaced with the Debian/Jessie distribution used on the Intel Edison.
\subsection{Fog Computing: Feature Extraction on Fog devices}
The fog devices, the Intel Edison and Raspberry Pi, were both
configured to run the same Debian/Jessie i386 distribution. Once the distribution was setup, both devices installed the same version of Octave 3.8.2-4, along with the additional packages required to perform the processing required by our algorithms. We also ensured that both gateway devices tracking system performance using the same tools. The tools we used included the Linux program \textit{top} and the Octave function \textit{Profiler}. The \textit{top} program provided real-time insights into CPU Load, Memory Usage, and run-times for processes or threads being managed by the Linux kernel. This was used later to provide use with benchmarking for the system overall. The Octave function \textit{Profiler} provided insights into the
run-times for each of section of the algorithm. This was used later determine which parts of the algorithm required more time to complete.
\subsection{Benchmarking and Program Setup}
The gateway devices where remotely logged into via the SSH protocol. From here we ran the same benchmarking scripts for both devices. The scripts would start Octave and load it with the data and use-case based algorithm, while top was started in parallel. The script searched
top for the process ID (PID) for this new instance of Octave. Once determined it would extract all the information top provided about the systems performance and the load imposed on the system by this instance of Octave. The extracted information was logged into a csv file and saved for analysis after the algorithm ran its course. Once the instance of Octave was ready to run the algorithm it started the Profiler function in the background. At the conclusion of the algorithm the Profilers set of data was stored into a .mat file for later analysis.
\subsection{Bandwidth~\& Data Reduction}
%
We conducted an experiment to measure the percentage by which Fog could reduce the data by processing the audio files using proposed Fog architecture. In our previous studies~\cite{dubey2015echowear}, we developed a clinical speech processing chain (CLIP), a series of filtering operations applied on the speech data for computing the clinical features such as loudness and fundamental frequency. We incorporated several new features in present chapter in addition to loudness and fundamental frequency used in~\cite{dubey2015echowear}.  We took 20 audio files and processed them with two methods;

\begin{enumerate}
\item Conventional method of compressing the files using GNU zip~\cite{gnu} and sending them to the cloud server for further processing ;

\item Extracting the clinical features on the fog computer (proposed Fog architecture).
\end{enumerate}
Table~\ref{table3} lists the performance of Fog computer with respect to computation of clinical features. Figure~\ref{fig_data_reduce} shows the percentage reduction in data size achieved by clinical speech processing and GNU zip compression. We can see that there is huge gains by processing data on Fog computer and sending only the features to cloud as compared to sending the original files to the cloud. 
\subsection{Engineering Perspectives}
Charging the wearables such as smartwatches~\emph{etc.} and gateways such as (smartphones/tablets) was necessary at least once in a day. In case patients want to do exercise while being away from home, they need to carry the tablet along
with them. Patients were asked to do exercise in a quiet place where the noise is very low or negligible. The patients could wear the smartwatch all the time. The tablet and the smartwatch need to be within a range of 50 meters. The
speech recordings were saved with date and time stamp that helped in sorting and query-ing them in cloud database. The participants have the choice to switch-on the recording system using smartwatch when they want to perform their vocal and other exercises. Similar procedures for other wearables.
\subsection{Medical Data Analytics and Visualization}
The part (a) of Figure~\ref{fig_data_stats} shows the size of speech data collected from one of the patients for eight days. We can see that the least amount of data (24 MB) was collected on the first day. On later days, the data size had been increasing. The part (b) of Figure~\ref{fig_data_stats} shows the patients feedback on using the \textit{IoT PD} technology for facilitating the remote monitoring of their vocal and speech exercises. Five out of six participants of in-home trials express a pleasant experience in using it. One participants had problems in using it for the first week. This patient had severe movement disorders in addition to speech disorder that made it difficult to switch ON/OFF the smartwatch. One week later, we made a software update allowing easier mechanism for switching ON/OFF. After using the updated \textit{IoT PD}, the patient reported that it was easy to use it. Accounting one feedback out of six as neutral, we depict the user experience as the pie chart shown in Figure~\ref{fig_data_stats} (b).

\section{Practical Insights}

\subsection{Data Vs. Fog Data for Cloud Storage} 
Table~\ref{fog_storage} shows approximation on the cloud storage requirement when we compare the conventional model of raw data transfer with the presented Fog Data. It is clear that for the long-term continuous data including speech and ECG, Fog Data architecture reduces the storage requirements tremendously and ultimately cuts the storage and maintenance cost as well as power demand on the cloud. Moreover, the reduced storage reduces the complexity of Big Data Analytics.   

\begin{figure*}[!tb]
	\centering
	\includegraphics[width=490pt]{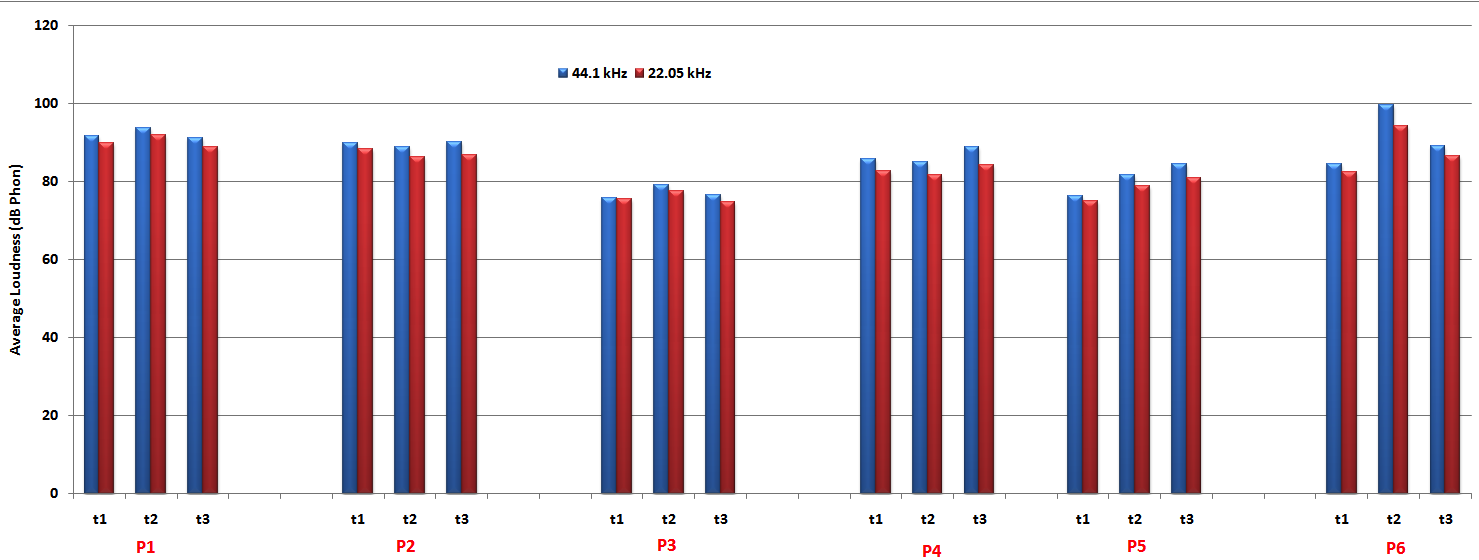}
	\caption{Showing effect of downsampling on loudness. We can see that by capturing pathological speech at lower sampling rate, we are still approximately at same loudness level. The lower sampling rate would lead to lower power consumption in battery-operated wearable devices.}
	\label{fig_22k_loud}
\end{figure*}

Figure~\ref{fig_22k_loud} shows the loudness computed by capturing pathological speech data at 44.1 kHz and downsampling it half the rate. It is clear that downsampling degrades the perceptual quality of pathological speech at the advantage of lower power consumption. This graph shows that if needed lower sampling rate can still be a useful in situations where power consumption on wearable devices is an issue.

\begin{table}[!tb]
\centering
\caption{Cloud storage requirement for 100 patients undergoing speech tele-therapy at home.}
	\begin{tabular}{c c c}
		\hline
		Time & Raw Data &  Fog Data \\
		\hline
		1 Day& 12 GB & 0.0012 GB\\ 
		\hline
		1 Week & 84 GB& 0.84 GB \\
		\hline
		1 Month & 360 GB & 3.6 GB\\
		\hline
		1 Year & 4079 GB & 43.8 GB \\
		\hline
		\label{fog_storage}
	\end{tabular}
\end{table}
\textbf{System Complexity:} Our experiences with Fog provide evidence that establishing an intelligent computing resource in remote settings where the patients were located was not only challenging in terms of hardware development and programming, but also required the interdependence of many
tools and libraries to build automated exchange of information among various elements of telemedicine. For example, Table~\ref{table4}  shows the various tools needed to bring autonomy, configurability, security, and
smart computing on the Intel Edison. We spent months to pursue a systematic survey of what was available and what was useful. Surveying the useful tools was time consuming yet rewarding.
\begin{table*}[!h]
	\small
	\centering
	\caption{Various resources required to develop Fog architecture.}
	\begin{tabular}{c c c}
		\hline
		Languages & Tools & Usage \\
		\hline
		MySQL & SequelPro, DataGrip & Run Queries and \ check database tables for testing purposes \\ \hline
		PHP & PhpStorm, Postman & \shortstack{Code the server to return\\ the data and information to the mobile application} \\
		\hline
		Python & PyCharm & \shortstack{Code the data Processing, storage, \\transmission, and interfacing
			with database} \\
		\hline
		Android & IntelliJ, Android Studio &\shortstack{ Coded the Client transfer \\ code and the complete mobile application} \\
		\hline
		All languages & Atom, Sublime Text2 & \shortstack{Editors which can \ code all languages}\\
		\hline
		\label{table4}
	\end{tabular}
\end{table*}
\subsection{Compatibility Issues}
There were countless instances when we had to find unconventional ways to
establish intelligence in Fog. For example, installing Praat python library on the Intel Edison was extremely difficult.

\subsection{Security~\& Privacy} 
In this work, we presented how the fog computer could be configured for computation and database access. We also touched upon Fog security from the authenticated access point-of-view. However, we believe that security needs can be addressed more rigorously since Fog  allows us to configure the fog computing node remotely and inject algorithms that could make the communication and storage more secure.   

\subsection{Challenges in using Fog computing for Telemedicine}
No system is perfect and fog computing is no exception. There are difficulties in deploying Fog architecture for telemedicine applications. Although the fog computing provides the data computation on the edge, reducing the data
significantly, the data becomes non-reversible when only analytics are communicated to the cloud. The fog has a limited storage space such that it can only store data for days or weeks, depending on the type of data. In our case, the data were audio files that could easily exceed the storage limit on the fog within a few days. An alternative is to create a query mechanism to access the data on fog when the clinicians want to listen to the audio files. Furthermore, since the raw data was not communicated to the cloud, there was no way to perform additional analysis in the cloud. In other words, it is necessary to ensure the reliability of the computational models used for analysis of the data before
they are injected into the fog computing resources.
\section{Conclusions}
We presented a multi-layer telemedicine architecture of the fog-assisted Medical Internet-of-things that was implemented on Intel Edison with layers for hardware, middleware (communication and software), and application (with security services). The Fog framework achieves intelligent gateway functions by processing audio files using signal processing algorithms such as psychoacoustic analysis to extract the clinical features; storage of raw data and features that are on-demand queryable by the cloud as well as the Fog interface. We also implemented Android apps for stakeholders such as patients, healthcare providers and administrators who require access to the backend database. This enabled speech-language pathologists (SLPs) to query the data showing daily progress of their patients. Our case study demonstrated that managing computations on Intel Edison (fog computer) reduces the data by 99\%; though less data reduction would occur if more features were analyzed. Our study also showed that it is possible to perform high-fidelity signal processing on the fog device to extract pathological speech features and communicate them to the cloud database. 

Moreover, the paper not only provides a high level understanding of the fog-based IoT system, but also provides details of how each layer was implemented including the tools and libraries used in the development. If implemented appropriately, Fog has a great potential to provide more autonomy and reliability in telemedicine applications driven by IoT. In future, we plan to deploy Fog in patient's homes. This will help us face operational challenges when the fog computer is located remotely in a different network.
\section*{Acknowledgments} 
Authors would like to thank the patients with Parkinson's disease for their co-operation during validation studies reported in this chapter. This work was supported by a grant (No: 20144261) from Rhode Island Foundation Medical Research and NSF grants CCF-1421823, CCF-1439011 and NSF CAREER CPS 1652538. Any opinions, findings, and conclusions or recommendations expressed in this material are those of the author(s) and do not necessarily reflect the views of the National Science Foundation or Rhode Island Foundation Medical Research. Authors would like to thank Alyssa Zisk for proofreading the manuscript. Authors would like to thank Manob Saikia, and Dr. Amir Mohammad Amiri for helpful discussions and suggestions for preparation of this chapter.    

\bibliographystyle{splncs03}
\bibliography{final_refs}
\end{document}